\documentclass[10pt]{article}
\pdfoutput=1
\usepackage{amsmath,amssymb,graphicx}
\usepackage{hyperref}
\usepackage[T1]{fontenc}
\usepackage[utopia]{mathdesign}
\usepackage[font=small,labelfont=bf]{caption}
\usepackage{longtable}
\usepackage{verbatim}
\usepackage{amsfonts}
\usepackage{cite}

\usepackage[usenames,dvipsnames]{xcolor}
\definecolor{darkerblue}{rgb}{0.2,0.2,0.5}

\usepackage{afterpage}

\renewcommand{\arraystretch}{1.25}
\newcommand{\bear}{\begin{array}}
\newcommand{\ear}{\end{array}}

\newcommand{\beq}{\begin{eqnarray}}
\newcommand{\eeq}{\end{eqnarray}}
\newcommand{\beqa}{\begin{eqnarray}}
\newcommand{\eeqa}{\end{eqnarray}}

\def\OMIT#1{{}}
\newcommand{\lsim}{\mathrel{\rlap{\lower4pt\hbox{\hskip1pt$\sim$}}
    \raise1pt\hbox{$<$}}}         
\newcommand{\gsim}{\mathrel{\rlap{\lower4pt\hbox{\hskip1pt$\sim$}}
    \raise1pt\hbox{$>$}}}         

\usepackage{titlesec}

\titleformat{\section}
{\color{darkerblue}\normalfont\Large\bfseries}
{\color{darkerblue}\thesection}{1em}{}
\titleformat{\subsection}
{\color{darkerblue}\normalfont\large\bfseries}
{\color{darkerblue}\thesubsection}{1em}{}

\title{\bf \color{darkerblue} A New Look at Higgs Constraints on Stops}

\author{JiJi Fan$^{a}$ and Matthew Reece$^{b}$\\
{\em $^a$ Department of Physics, Syracuse University, Syracuse, NY, 13210, USA} \\
{\em $^b$ Department of Physics, Harvard University, Cambridge, MA 02138, USA}}

\begin{document}
\maketitle

\begin{abstract}
We present a simple new way to visualize the constraints of Higgs coupling measurements on light stops in natural SUSY scenarios beyond the MSSM, which works directly in the plane of stop mass eigenvalues (with no need to make assumptions about mixing angles). For given stop mass eigenvalues, the smallest value of $X_t$ that can bring the correction to the $h \to gg$ and $h\to \gamma\gamma$ couplings into agreement with data is computed. Requiring that this $X_t$ is consistent---i.e. that the chosen mass eigenvalues can be the outcome of diagonalizing a matrix with a given off-diagonal term---rules out the possibility that both stops have a mass below $\approx 400$ GeV. Requiring that $X_t$ is not fine-tuned for agreement with the data shows that neither stop can be lighter than $\approx 100$ GeV. These constraints are interesting because, unlike direct searches, they apply no matter how stops decay, and suggest a minimum electroweak fine-tuning of between a factor of 5 and 10. We show that a multi-parameter fit can slightly weaken this conclusion by allowing a large Higgs coupling to $b$-quarks, but only if a second Higgs boson is within reach of experiment. Certain models, like $R$-symmetric models with Dirac gauginos, are much more strongly constrained because they predict negligible $X_t$. We illustrate how the constraints will evolve given precise measurements at future colliders (HL-LHC, ILC, and TLEP), and comment on the more difficult case of Folded Supersymmetry. 
\end{abstract}

\section{Introduction}
\label{sec:intro}
The LHC experiments have undertaken a serious effort to discover ``natural supersymmetry,'' focusing on higgsinos, stops, and gluinos as the tree-level, one- and two-loop harbingers of naturalness~\cite{Dimopoulos:1995mi, Cohen:1996vb}. Direct experimental constraints have already ruled out many such natural scenarios~\cite{Kribs:2013lua,Arvanitaki:2013yja,Krizka:2012ah,Evans:2013jna,Han:2013kga}. The Higgs mass itself constrains the MSSM to an unnatural region of parameter space, so we will always have in mind theories beyond the MSSM where new interactions lift the Higgs mass. On the other hand, model builders can design scenarios that evade direct searches~\cite{Barbier:2004ez, Burdman:2006tz,Fan:2011yu,Fan:2012jf,Alves:2013wra}, prolonging a definite conclusion until a lengthy game of cat-and-mouse is completed. Another route to constrain naturalness is to rely on the fact that any natural Higgs is {\em not} the Standard Model Higgs, and so modifications of Higgs properties must eventually show up in a natural theory~\cite{Arvanitaki:2011ck,Blum:2012ii,Gupta:2012fy,Craig:2013xia,Farina:2013ssa,Kribs:2013lua}. In particular, loops of light stops modify the Higgs couplings to gluons and photons, so measurements of Higgs properties give a bound on stops that is independent of their decay mode and so applies even if they have hidden from direct searches. Our goal in this paper is to provide a perspective on these constraints that clarifies the physics and makes it easier to understand what Higgs measurements are telling us so far about fine-tuning.

Our main innovation, relative to earlier studies of Higgs constraints on stops, is to provide a two-dimensional plot that clearly shows how the measured Higgs properties constrain the amount of fine-tuning arising in the stop sector. The stop mass matrix depends on three parameters: the two soft mass terms $m_{Q_3}^2$ and $m_{u_3}^2$ and the mixing parameter $X_t = A_t - \mu/\tan\beta$. As a result, prior studies typically plot constraints in a variety of two-dimensional slices of the general three-dimensional parameter space, for instance by fixing the mixing angle relating the basis of left- and right-handed stops to the mass eigenbasis.

The insight we exploit is the following: the most interesting parameters (for instance for designing collider searches) are the {\em physical} mass eigenvalues $m_{{\tilde t}_{1,2}}$. So we plot the bounds in the plane of physical masses. For any given pair of physical mass eigenvalues, there is a {\em largest} consistent choice of $X_t$. As we will discuss, this is a straightforward consequence of diagonalizing the $2\times 2$ stop mass matrix. On the other hand, for light stops one finds a large positive correction to the gluon fusion cross section unless there is a {\em minimum} $X_t$ that allows the correction to be canceled. For a given pair of physical stop mass eigenvalues, if the minimum $X_t$ consistent with the measured cross section is larger than the maximum $X_t$ that is theoretically consistent, the mass point is ruled out {\em for all values of the mixing angle}. We can supplement this with a region that is excluded by the need to fine-tune the value of $X_t$ precisely to allow a fit to the data. These arguments allow us to make a readily comprehensible two-dimensional plot that shows the excluded range of stop masses and (with certain assumptions we will specify) a minimum amount of fine-tuning implied solely by Higgs measurements.

The importance of this insight is that it gives a straightforward way to go from measured Higgs data to a claim about fine-tuning in supersymmetric theories. The argument is quite robust in the sense that we constrain stops no matter how they decay, so theories with $R$-parity violation or ``stealthy'' decay modes are not exempt from the argument. Loopholes could come from special UV physics that ameliorates the low-energy estimate of fine-tuning, but we expect that this rarely alters the conclusion in a qualitative way. We discuss how future precise measurements of Higgs properties will allow our argument to be refined.

Previous papers that have examined constraints on the stop sector based on LHC Higgs data include~\cite{Arvanitaki:2011ck, Carena:2011aa,Carmi:2012yp,Cohen:2012zza,Curtin:2012aa,Giardino:2012ww,Carmi:2012in,Carena:2012np,Espinosa:2012in,D'Agnolo:2012mj,Falkowski:2013dza,Giardino:2013bma,Farina:2013ssa,Gori:2013mia}. Authors of these papers made a variety of choices for how to plot the constraints. For instance, some fix $X_t$ or the mixing angle $\theta_{\tilde t}$ and plot in the plane of stop mass eigenvalues $\left(m_{{\tilde t}_1}, m_{{\tilde t}_2}\right)$; or assume the left-handed stops are decoupled and plot only constraints on right-handed stops (relevant for electroweak baryogenesis); or plot in the $\left(m_{{\tilde t}_1}, m_{{\tilde t}_2} - m_{{\tilde t}_1}\right)$ plane for specific choices of mixing angle $\theta_{\tilde t}$. Other studies impose that stops lift the Higgs mass to 125 GeV in the MSSM (see, for instance, ref.~\cite{Barger:2012hr,Delgado:2012eu,Carena:2013iba}) and so consider only a slice of the full parameter space. Before the LHC had delivered any data, Ref.~\cite{Dermisek:2007fi} considered how the gluon fusion rate at the LHC could be used to probe the plane of average stop mass and $X_t/m_{\tilde t}$. Constraints on large values of the mixing $X_t$ from vacuum instabilities have also received a great deal of attention~\cite{Kusenko:1996jn,Reece:2012gi,Camargo-Molina:2013sta,Chowdhury:2013dka,Blinov:2013fta}. All of these contributions are useful and have some overlap with our work, but our visualization of the constraints is a more effective way to extract the bottom line: what do measured Higgs properties tell us about allowed stop masses?

In Section~\ref{sec:basic}, we explain the idea of how we exclude regions in stop mass space in more detail. We present the actual constraints in Section~\ref{sec:constraints}. We begin in Sec.~\ref{sec:case1} with a fit where the only corrections arise from stops running in loops. We neglect, for example, the possibility of additional loop corrections from non-MSSM states. This gives a simple result that we can summarize by saying that at least one stop should be heavier than $\approx 400$ GeV and neither stop can be lighter than $\approx 100$ GeV. We then consider variations: in Sec.~\ref{sec:case2}, we also allow the Higgs coupling to $b$-quarks to vary. This is the leading correction expected at large $\tan \beta$ when the 125 GeV Higgs mixes with a heavier Higgs boson. We find that the limit can be slightly weaker if the heavier Higgs mass is below 500 GeV (and thus in reach of experiments in the near future). We then consider a four-parameter fit in Sec.~\ref{sec:case3}, finding that if $\tan \beta \approx 1$ and the Higgs coupling to $b$-quarks is enhanced, much more parameter space for stops opens up. This is a further important motivation for searching for heavy Higgs bosons: tighter constraints on their masses turn into tighter constraints on stops from the Higgs fit.  (Further naturalness arguments for searching for the heavy Higgs bosons may be found in ref.~\cite{Gherghetta:2014xea}.) We conclude our look at stop constraints in Sec.~\ref{sec:future} by showing that the ILC or TLEP would significantly extend the excluded region of stop masses by performing accurate measurements of Higgs couplings. In Section~\ref{sec:folded}, we look at the more exotic model of Folded Supersymmetry, where the constraints are currently very weak but TLEP would still give a nontrivial exclusion. We conclude in Section~\ref{sec:discussion} with a discussion of the implication of our results for fine-tuning.

Appendix~\ref{app:AtermRG} shows that our results become stronger in models where $A_t$ is generated radiatively from the gluino mass, because in that case a minimum $A_t$ from data translates into a minimum gluino mass, which contributes to electroweak fine tuning at two-loop order. Two further appendices give technical details: the data used in our fits are summarized in Appendix~\ref{app:data}, and our results are compared to those of the HiggsSignals code in Appendix~\ref{app:HScompare}.

\section{Basic Idea}
\label{sec:basic}

Before beginning, we comment on notation: we define the modifications to the Higgs couplings to SM particles as 
\beq\label{eq:r}
r_{i}\equiv\frac{c_{hii}}{c^{\rm SM}_{hii}}, 
\eeq
with $c$'s denoting couplings and $i = t, V, G, \gamma, b, \tau$ standing for top, massive vector gauge bosons, gluon, photon, bottom and tau respectively.\footnote{We take further $r_W=r_Z=r_V$ although this may have exceptions~\cite{Farina:2012ea}.}

The stop mass-squared matrix, in the gauge eigenstate basis $(\tilde{t}_L, \tilde{t}_R)$, is given by 
\[ \left( \begin{array}{cc}
m_{Q_3}^2+m_t^2 + \Delta_{\tilde{u}_L} & m_t X_t  \\
m_t X_t^* & m_{U_3}^2+m_t^2+\Delta_{\tilde{u}_R} \end{array} \right),\]
where $m_{Q_3}^2, m_{U_3}^2$ are the soft mass squared of left- and right- handed stops respectively and the stop mixing term $X_t = A_t - \mu/\tan\beta$. For simplicity, we will neglect possible phases in the stop mass matrix. $\Delta_{\tilde{u}_L}=\left(\frac{1}{2} - \frac{2}{3}\sin^2\theta_W\right) \cos(2\beta) m_Z^2$ and $\Delta_{\tilde{u}_R}=\left(\frac{2}{3} \sin^2\theta_W\right)\cos(2\beta)m_Z^2$ originate from the $D$-term quartic interactions and are $\ll m_t^2$.

It is easy to see that the off-diagonal stop mixing terms always split the two mass eigenstates. 
More specifically, the splitting between two physical masses squared can be expressed in terms of the mass parameters as 
\beq
\left |m_{\tilde{t}_1}^2-m_{\tilde{t}_2}^2\right| = \sqrt{(m_{Q_3}^2+\Delta_{\tilde{u}_L} -m_{U_3}^2-\Delta_{\tilde{u}_R})^2+4 m_t^2 X_t^2},
\eeq
where the first term in the square root comes from the difference in the diagonal mass terms while the second one comes from the off-diagonal mass term. Thus for fixed physical stop masses, the maximally allowed $X_t$ is given by 
\beq\label{eq:Xtmax}
\left|X_t^{\rm max} \right|= \frac{\left|m_{\tilde t_1}^2 - m_{\tilde t_2}^2 \right|}{2 m_t},
\eeq 
which is only achieved when the diagonal mass terms are equal. In particular, two mass degenerate stops correspond to $X_t = 0$.

As is well known, stop loops could modify the Higgs coupling to gluons, of which the leading order contribution could be computed easily via the low energy Higgs theorem~\cite{Ellis:1975ap, Shifman:1979eb}
\beq\label{eq:rG}
r_G^{\tilde t} \equiv \frac{c_{hgg}^{\tilde t}}{c_{hgg}^{\rm SM}}
\approx \frac{1}{4} \left(\frac{m_t^2}{m_{\tilde{t}_1}^2}+\frac{m_t^2}{m_{\tilde{t}_2}^2}-\frac{m_t^2X_t^2}{m_{\tilde{t}_1}^2m_{\tilde{t}_2}^2}\right), \quad {\rm stop \, contribution,}
\eeq
where we neglect $D$-terms. This expression is valid for $m_{{\tilde t}_{1,2}} \gsim m_h/2$, which we will assume. Exotic scenarios where lighter stops could have evaded detection would also predict a large Higgs decay rate to stops, so it is safe to dismiss the possibility. One can see that without mixing ($X_t \approx 0$) light stops could give a considerable positive contribution to $r_G^{\tilde{t}}$. If it exceeds the upper bound allowed by the Higgs coupling measurements, there has to be a cancelation between the first two positive terms and the last negative term from stop mixing. The low-energy theorem asserts that the loop correction from a particle with mass $M(v)$ is $\propto \partial \log M^2(v)/\partial \log v$; the mixing contributes negatively because a larger Higgs vev would mean a larger off-diagonal term and would decrease the lightest stop mass. Thus for light stops to be consistent with the Higgs coupling data, $X_t$ has to be larger than 
\beq \label{eq:Xtmin}
\left|X_t^{\rm min} \right|=\frac{\sqrt{m_t^2(m_{\tilde t_1}^2+m_{\tilde t_2}^2)-4\left(r_G^{\tilde t}\right)^{\rm fit;max}m_{\tilde t_1}^2m_{\tilde t_2}^2}}{m_t},
\eeq
where $\left(r_G^{\tilde t}\right)^{\rm fit;max}$ is the upper end of the experimental allowed range from a fit. We will describe the procedure of the fit in the next section. This formula is only valid when the quantity in the square root in Eq.~\ref{eq:Xtmin} is positive; otherwise, there is no constraint. 

For given $(m_{\tilde t_1}, m_{\tilde t_2})$, if $\left|X_t^{\rm max} \right| $ in Eq.~\ref{eq:Xtmax} allowed by the physical masses is smaller than $\left|X_t^{\rm min} \right|$ in Eq.~\ref{eq:Xtmin} allowed by the Higgs coupling, this point in the parameter space is inconsistent with the Higgs coupling measurements. This constraint will be strongest along the mass degenerate line, $m_{\tilde t_1} = m_{\tilde t_2}$, where the physical masses only allow zero mixings.  Although equation~\ref{eq:rG} is not valid for stop masses less than about half the Higgs mass (at which point the low energy theorem must be modified by including the appropriate loop function), considering its $m_{{\tilde t}_1} \to 0$ limit is instructive for understanding the constraints. In this limit, the only way to cancel the large correction is to set $X_t = m_{{\tilde t}_2}$. In that case, we can see from eq.~\ref{eq:Xtmax} that the excluded region would be $m_{{\tilde t}_2} < 2 m_t$. As we will see, our exclusion plots have this feature: the curves (extrapolated to smaller stop masses) are ``anchored'' at the points $(0,2 m_t)$ and $(2 m_t, 0)$ on the axes, and extend away from these points along the diagonal. An alternative argument reaches parameter space farther from the diagonal: for some choices of stop masses, $\left|X_t^{\rm min}\right|$ is not larger than $\left|X_t^{\rm max}\right|$, but the degree of cancelation required in eq.~\ref{eq:rG} to fit the data is so large that the point requires a high amount of fine-tuning to fit the data. At present the region excluded by this tuning argument is small, but it will become more important with future precise measurements. In Sec.~\ref{sec:constraints}, we will follow this basic idea and present the constraints from Higgs coupling measurements in the stop mass plane in three different cases. 

It is also worth noting that in certain models, $X_t$ is expected to be very small. This is, in particular, the case for $R$-symmetric models. These have received a great deal of recent attention because Dirac gluinos ameliorate the ``second-order naturalness problem'' of the gluino mass lifting the stop mass through RGEs~\cite{Fox:2002bu,Nelson:2002ca,Kribs:2007ac,Kribs:2012gx} (though see~\cite{Arvanitaki:2013yja,Csaki:2013fla} for a less sanguine take). In such models, the experimental constraint from Higgs data on stop masses will be much stronger, because we no longer have the freedom to cancel a positive contribution against a negative one from mixing. We will illustrate this stronger constraint on $R$-symmetric models below.

\subsection{Global Fit of Higgs Couplings}
\label{sec:fit}
The main tool we employ to extract constraints on natural SUSY is a global fit of Higgs data. Right now there are five major final states in Higgs searches: $h \to \gamma \gamma, h \to ZZ^{*} \to 4 \ell, h \to WW^{*}, h \to b\bar{b}, h \to \tau \bar{\tau}$. In each channel, the ratio of the expectation of a BSM scenario compared to the SM predication can be expressed as a function of $r_i$: 
\beq
\mu_f=\left(\sum_i r_i^2 \xi_i\right)\left( \frac{ r_f^2}{\sum_j r_j^2 \,{\rm Br} (h \to jj)}\right),
\eeq
where the first summation is over all Higgs production modes denoted by $i$'s while the second is over all Higgs decay channels denoted by $j$'s. $\xi_i$ stands for the fraction of the signal events contributed by a specific production channel $i$ and identifies the appropriate weight of each coupling in a particular channels' production rescaling. Then to assess the compatibility of a point in the parameter space with the Higgs data, we construct a $\chi^2$ function
\beq
\chi^2 = \sum_f \frac{\left(\mu_f - \mu_f^{obs}\right)^2}{\sigma_f^2},
\eeq
where $\mu_f^{obs}$ is the central value of the observed data in channel $f$ and $\sigma_f$ denotes the associated $1 \sigma$ error bar. The data we will use in our work is listed in Appendix~\ref{app:data}.

\section{Constraints on Natural Stops}
\label{sec:constraints}
In this section we follow the procedure in Sec.~\ref{sec:basic} and explore the implications of Higgs fit results for natural stops. We consider three cases with increasing number of parameters that parametrize the modifications of Higgs couplings.

\subsection{Case 1: Stop Loops, No Higgs Mixing}
\label{sec:case1}

First we consider the simplest case with only one source of Higgs coupling modification. In this scenario, the contributions to the Higgs couplings from the light Higgs mixing with other scalars are negligible. This could naturally arise, e.g., in the decoupling limit when the other Higgses are heavy. We also assume that the chargino contribution to the Higgs diphoton coupling is negligible, which is true for $\tan\beta \gtrsim 3$.\footnote{Exceptions see~\cite{Batell:2013bka}.} Consequently in this case we only have one free parameter $r_G^{\tilde{t}}$, which is already discussed in Sec. 2. 
The stops' contribution to the Higgs diphoton coupling is smaller and anti-correlated with $r_G^{\tilde t}$,
\beq\label{eq:gamG} r_\gamma^{\tilde t}  \equiv \frac{c_{h\gamma\gamma}^{\tilde t}}{c_{h\gamma\gamma}^{\rm SM}}=\frac{\mathcal{A}^\gamma_{\tilde t}}{\left(\mathcal{A}^\gamma_W+\mathcal{A}^\gamma_t\right)^{\rm SM}}
\approx -0.28 r_G^{\tilde t},\eeq
using $\mathcal{A}^\gamma_W\approx8.33$ and $\mathcal{A}^\gamma_t\approx-1.84$, the amplitudes of $h \to \gamma\gamma$ in the SM, valid for $m_h=125$ GeV. 

\begin{figure}[!h]\begin{center}
\includegraphics[width=0.45\textwidth]{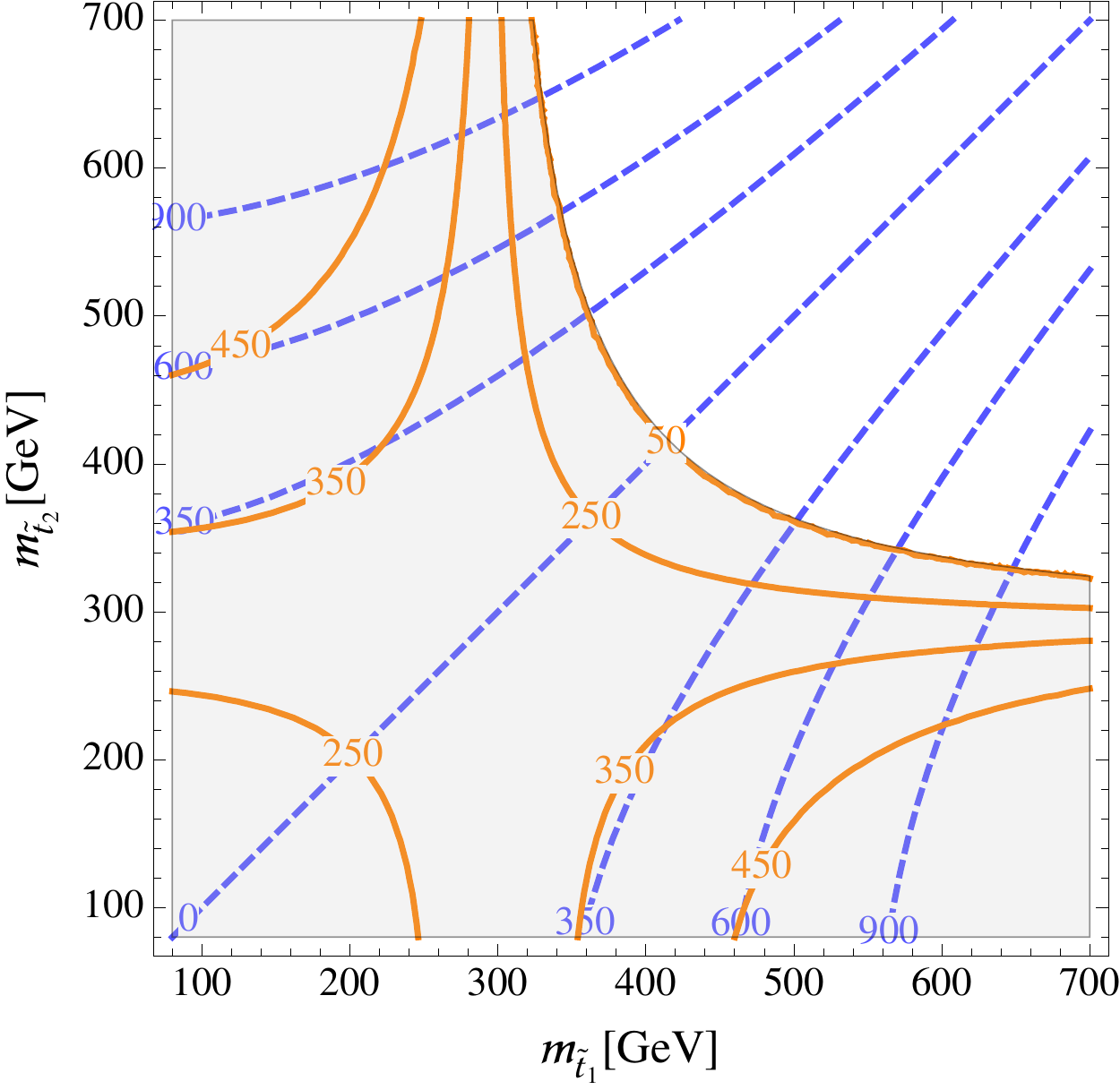}
\end{center}
\caption{Illustration of the principle behind our exclusion plots. The blue dashed contours are the largest allowed mixing parameters for given stop mass eigenvalues, $\left|X^{\rm max}_t\right|$ (as in eq.~\ref{eq:Xtmax}). The orange solid contours are the minimum mixing $\left|X^{\rm min}_t\right|$ required to fit the data at 2$\sigma$, as in eq.~\ref{eq:Xtmin}, under the hypothesis that only stop loops modify Higgs couplings. In the case of models with an $R$-symmetry where $X_t = 0$, the entire shaded gray region is excluded at 2$\sigma$ by the data. In more general models, we display the exclusion below.}
\label{fig:methodillustrate}
\end{figure}%

\begin{figure}[!h]\begin{center}
\includegraphics[width=0.45\textwidth]{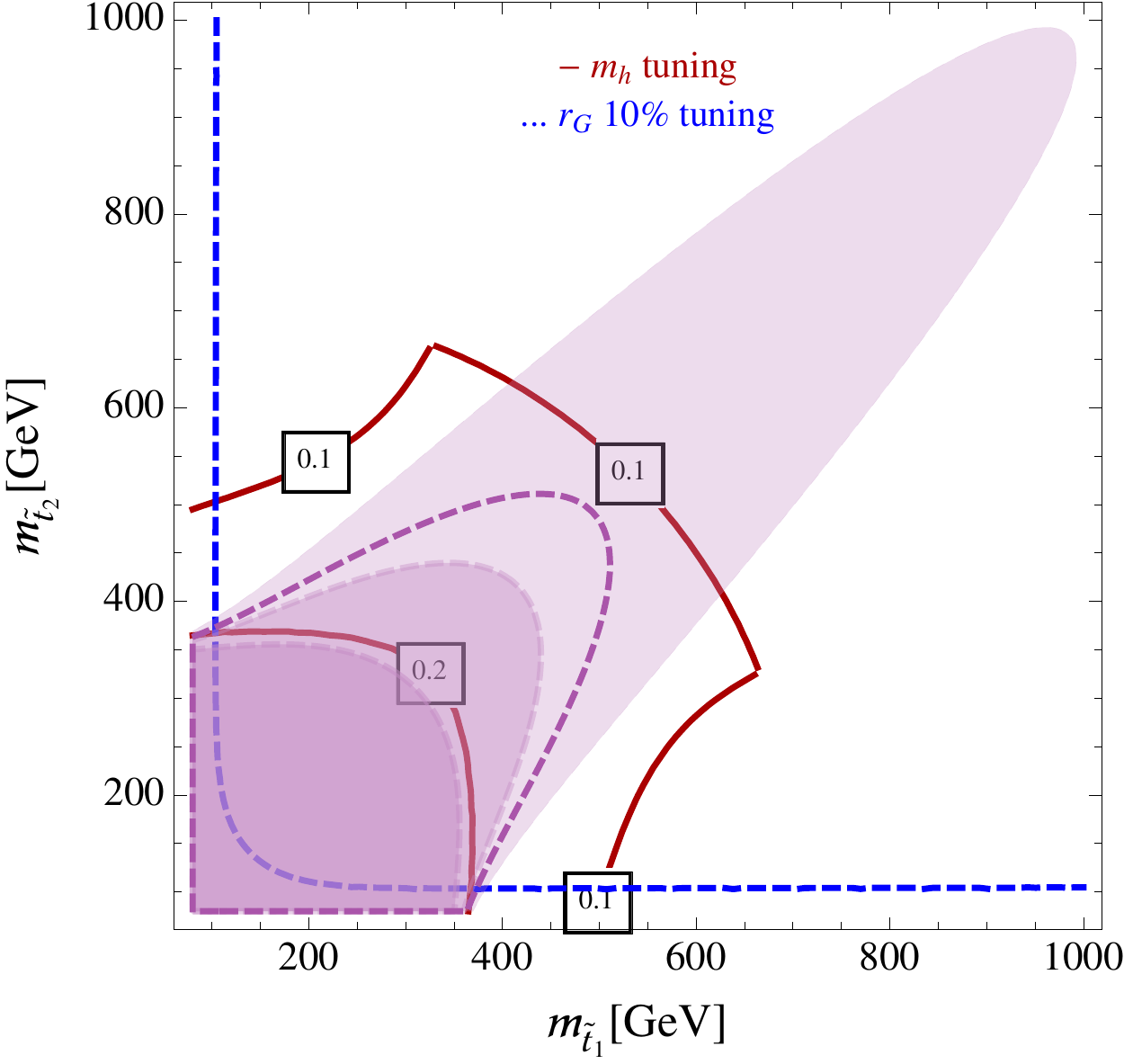}
\end{center}
\caption{Assuming no other contributions to Higgs digluon coupling $r_G$ other than stops', region of natural stop that has been ruled out by Higgs coupling measurements. The three shaded purple regions, from darkest to lightest, are excluded at $3 \sigma$ (99.73\%) level; $2 \sigma$ (95.45\%) level; and $1 \sigma$ (68.27\%) level. The dashed purple line is the boundary of the region excluded at 90\% CL. The red solid lines are contours of Higgs mass fine-tuning assuming $\Lambda = 30$ TeV, $\mu = -200$ GeV and $\tan\beta =10$. We have evaluated the tuning with $X_t = X_t^{\rm min}$, the smallest mixing allowed by the data at $2\sigma$ for a given pair of masses. The blue dashed line is a contour of 10\% fine-tuning associated with $r_G^{\tilde{t}}$. }
\label{fig:higgscoupling}
\end{figure}

As discussed in Sec.~\ref{sec:basic}, for a given point in the $(m_{\tilde t_1}, m_{\tilde t_2})$ plane, if $\left|X_t^{\rm max} \right| $ in Eq.~\ref{eq:Xtmax} allowed by the physical masses is smaller than $\left|X_t^{\rm min} \right|$ in Eq.~\ref{eq:Xtmin} allowed by the Higgs coupling, this point is excluded by the Higgs coupling measurements. We illustrate this principle in Fig.~\ref{fig:methodillustrate}, which shows contours of $\left|X_t^{\rm min,max}\right|$. The shaded region in Fig.~\ref{fig:methodillustrate} is ruled out in models where $X_t \approx 0$, e.g. $R$-symmetric theories. The excluded region allowing for nonzero $X_t$ is demonstrated in Fig.~\ref{fig:higgscoupling}.

In Fig.~\ref{fig:higgscoupling}, we also plot the Higgs mass fine-tuning, which is defined as~\cite{Kitano:2006gv,Perelstein:2007nx}
\beq\label{eq:Dz}
\left(\Delta_h^{-1}\right)_{\tilde t}=\left|\frac{2 \delta m_{H_u}^2}{m_h^2}\right|,\;\;\;\quad \delta m_{H_u}^2|_{\rm stop}&=&-\frac{3}{8\pi^2}y_t^2\left(m_{Q_3}^2+m_{U_3}^2+A_t^2\right)\log\left(\frac{\Lambda}{m_{\rm EW}}\right) \nonumber \\
&=&-\frac{3}{8\pi^2}y_t^2\left(m_{\tilde t_1}^2+m_{\tilde t_2}^2-2m_t^2+A_t^2\right)\log\left(\frac{\Lambda}{m_{\rm EW}}\right).
\eeq
Here $\Lambda$ is a scale characterizing mediation of SUSY breaking, while $m_{\rm EW}$ is the low scale at which running stops. We take $m_{\rm EW} = \max(\sqrt{m_{{\tilde t}_1} m_{{\tilde t}_2}}, m_h)$. In Fig.~\ref{fig:higgscoupling}, we take $A_t = \max\left(0,\left|X_t^{\rm min} \right|+\mu/\tan \beta\right)$ with the SUSY breaking mediation scale $\Lambda = 30$ TeV, $\mu = -200$ GeV and $\tan\beta =10$. The $\max$ here ensures that if the $\mu$-term alone is enough to provide $\left|X_t\right| > \left|X_t^{\rm min}\right|$, we set $A_t = 0$. Here $\left|X_t^{\rm min}\right|$ is taken to be the smallest value allowed at 2$\sigma$. We have deliberately chosen a very low mediation scale as well as a negative sign of $\mu$ relative to $A_t$ in order to draw conservative conclusions about the tuning measure. One could try to always generate $\left|X_t^{\rm min}\right|$ mostly from the $\mu/\tan\beta$ term, but this leads to tree-level tuning that is much worse than the loop-level tuning from $A_t$. To get the Higgs coupling within the allowed range of experiments, there could be a cancelation between contributions with opposite signs from the diagonal masses and mass mixings between two stops. Thus one could also define a fine-tuning measure associated with the Higgs coupling 
\beq
\left(\Delta_{G}^{-1}\right)_{\tilde t}=\left| \sum_i\left( \frac{\partial \log r_G^{\tilde{t}}}{\partial \log p_i} \right)^2\right|^{1/2},
\eeq
with the parameter set denoted by $p = (m_{Q_3}^2, m_{U_3}^2, X_t)$. In the limit $X_t^2 \approx m_{{\tilde t}_1}^2 + m_{{\tilde t}_2}^2$ where the coupling correction vanishes, this scales with the amount of tuning in the sense that
\beq
\left(\Delta_{G}^{-1}\right)_{\tilde t} \sim \left|\frac{X_t^2}{m_{{\tilde t}_1}^2 + m_{{\tilde t}_2}^2-X_t^2}\right|.
\eeq
So far the precision level of Higgs coupling measurements is still low, thus the fine-tuning of Higgs couplings is not very large in general. In Fig.~\ref{fig:higgscoupling}, we plot the boundary corresponding to 10\% fine-tuning in Higgs coupling, which excludes the possibility that even one stop is below about 100 GeV. (This is, essentially, the same observation that was made in the context of electroweak baryogenesis in Refs.~\cite{Cohen:2012zza,Curtin:2012aa}.) 
We also considered contributions from light stops to electroweak precision observables, in particular, the $\rho$ parameter, but the constraints there are much weaker compared to those from current Higgs coupling measurements. 

From Fig.~\ref{fig:higgscoupling}, we see that regions with both stops lighter than about 400 GeV is excluded by the Higgs coupling measurements at $2 \sigma$ (95.45 \%) C.L. Along the diagonal line where both stops are degenerate in mass, the constraint gets stronger and extends to 450 GeV. 
In general, although one could construct clever natural models where stops with different decaying topologies could evade the current collider searches, the Higgs coupling measurements provide a powerful indirect probe independent of the stop decays. One can also see that at 3$\sigma$ level, 20\% fine-tuning of Higgs mass, meaning that loop-level contribution to the Higgs mass is about the same as the tree-level Higgs mass, is inconsistent with the Higgs coupling measurements. A 10\% fine-tuning is still compatible with the data at 90\% confidence level, although a substantial portion of the parameter space with less than 10\% tuning is already ruled out.

\subsection{Case 2: Higgs Mixing Effect at Large $\tan\beta$}
\label{sec:case2}
Now we consider a slightly more complicated case with two parameters parametrizing the new physics contributions to Higgs couplings. In the scenario, besides the stops' contribution to Higgs digluon (and correlated diphoton) coupling, the Higgs mixing effects could be parametrized by a single parameter $r_b$, the ratio of bottom Yukawa coupling in the new scenario vs. the SM one. This is the case when $\tan \beta$ is large, i.e., $\tan \beta \gtrsim 3$. To see this, one recalls that in 2HDM at tree level, 
 $r_b=r_\tau$, and 
\beq\label{eq:rs2hdm}
r_b=\frac{vc_{hb\bar b}}{m_b}=-\frac{\sin\alpha}{\cos\beta},\;\;\;r_t=\frac{vc_{ht\bar t}}{m_t}=\frac{\cos\alpha}{\sin\beta},\;\;\;r_V=\frac{vc_{hVV}}{2m_V^2}=\sin\left(\beta-\alpha\right),
\eeq
implying the inequalities
\beq r_b^2\leq\tan^2\beta+1,\;\;\;r_t^2\leq\frac{1}{\tan^2\beta}+1,\;\;\;r_V^2\leq1.\eeq
So in general there are two independent parameters to describe $r_b,\,r_t$ and $r_V$. We choose these parameters to be $\tan\beta$ and $r_b$. With this choice we write
\beq\label{eq:parr} r_t=\sqrt{1-\frac{r_b^2-1}{\tan^2\beta}},\;\;\;\;r_V=\frac{\tan\beta}{1+\tan^2\beta}\left(\frac{r_b}{\tan\beta}+\sqrt{1+\tan^2\beta-r_b^2}\right),\eeq
valid for all $\tan\beta$. For large $\tan\beta$, $r_t \approx r_V \approx 1$, thus effectively we are left with only one parameter $r_b$. 

\begin{figure}[!h]\begin{center}
\includegraphics[width=0.43\textwidth]{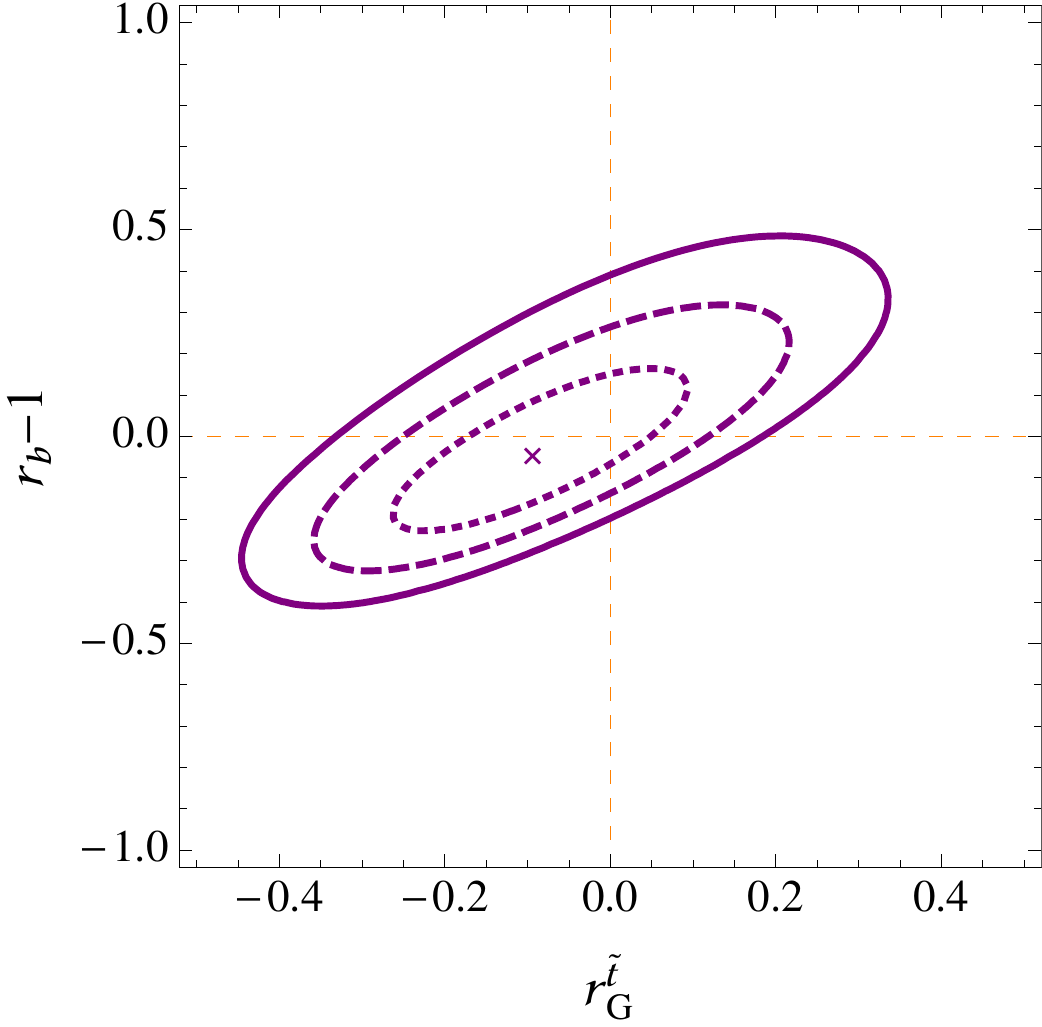}  \quad \includegraphics[width=0.45\textwidth]{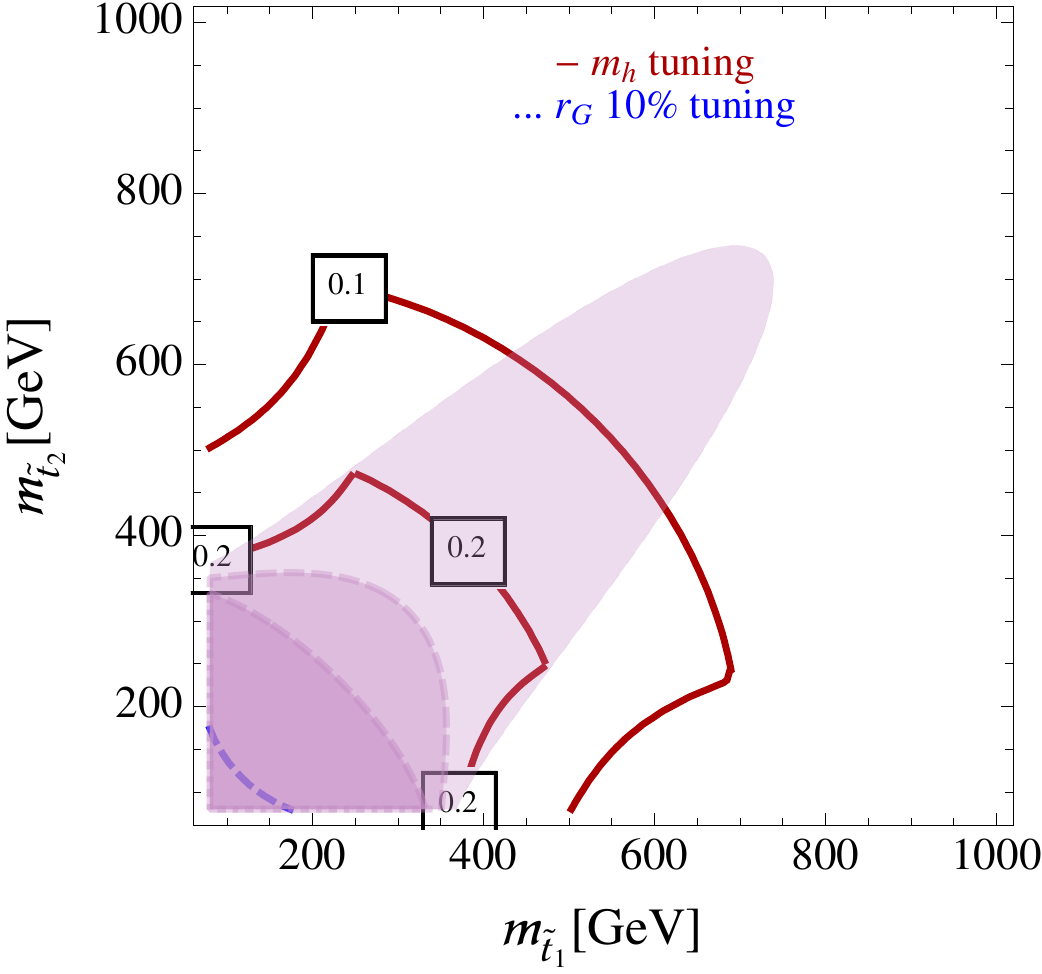}
\end{center}
\caption{Left: global fit in the $(r_G^{\tilde t}, r_b-1)$ plane. $\times$ denotes the best fit point. The dotted, dashed, and solid purple contours denote the boundaries of the allowed region at $1 \sigma$, $2 \sigma$, $3 \sigma$ C.L. Right panel: assuming no other contributions to Higgs digluon coupling $r_G$ other than stops', region of natural stop that has been ruled out by Higgs coupling measurements with varying $r_b$. The three shaded purple regions, from darkest to lightest, are excluded at $3 \sigma$ (99.73\%) level; $2 \sigma$ (95.45\%) level; and $1 \sigma$ (68.27\%) level.  The red solid lines: contours of Higgs mass fine-tuning assuming $\Lambda = 30$ TeV, $\mu = -200$ GeV and $\tan\beta =10$; blue dashed lines: contour of 10\% fine-tuning associated with $r_G^{\tilde{t}}$.}
\label{fig:fitp2}
\end{figure}

In the left panel of Figure~\ref{fig:fitp2}, we plot the boundaries of the allowed region at $1 \sigma$, $2 \sigma$, $3 \sigma$ C.L. in the $(r_G^{\tilde t}, r_b-1)$ plane from the global fit of the Higgs data. One could see that if the bottom Yukawa coupling is enhanced, the allowed $r_G^{\tilde t}$ also increases. 

We also perform a profile likelihood fit to map out the allowed region in the plane of the stops' physical masses. This is depicted in the right panel of Fig.~\ref{fig:fitp2}. One could see from the plot that the excluded regions shrink a bit compared to the case with only one parameter. Yet still region with two light stops below 350 GeV is excluded by the Higgs data at 2$\sigma$ level. 

\begin{figure}[!h]\begin{center}
\includegraphics[width=0.43\textwidth]{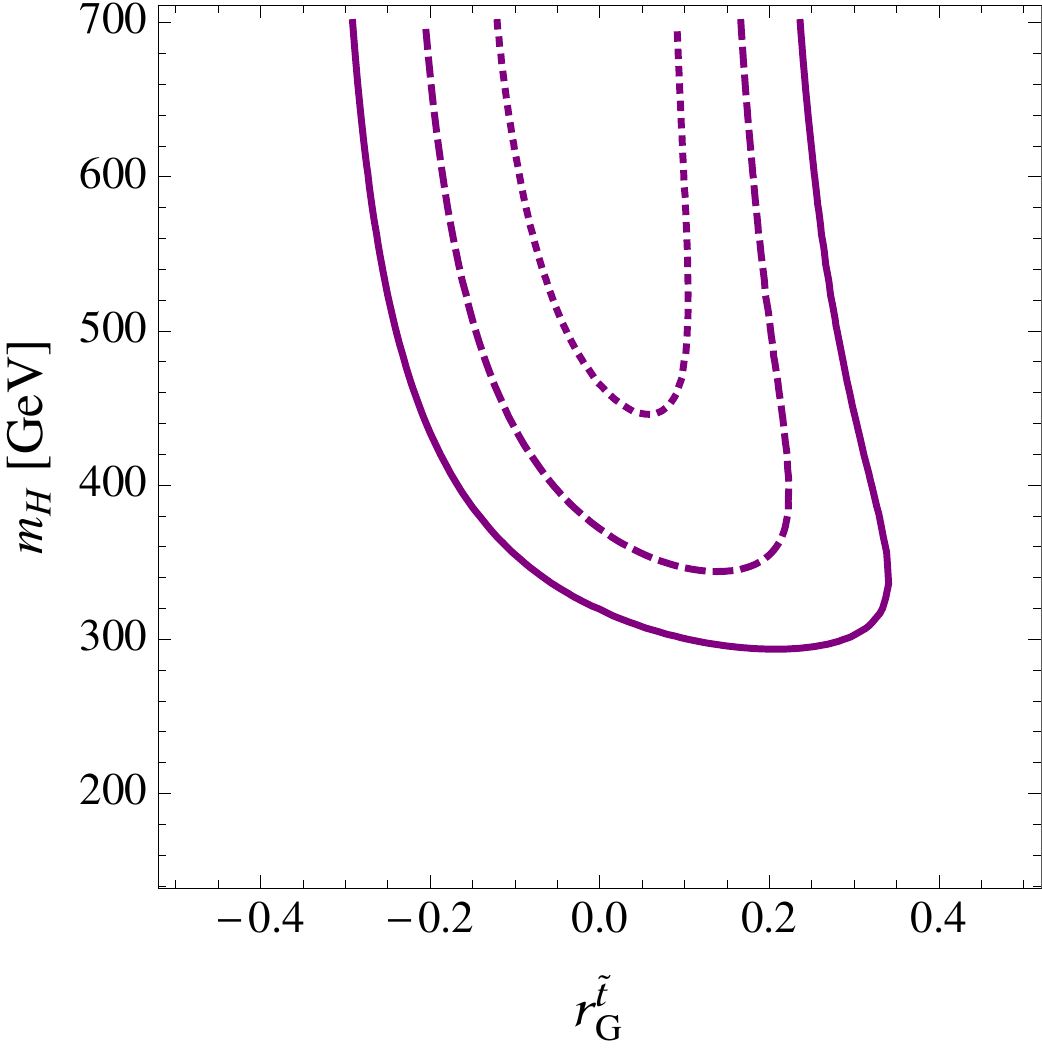}   \quad \includegraphics[width=0.43\textwidth]{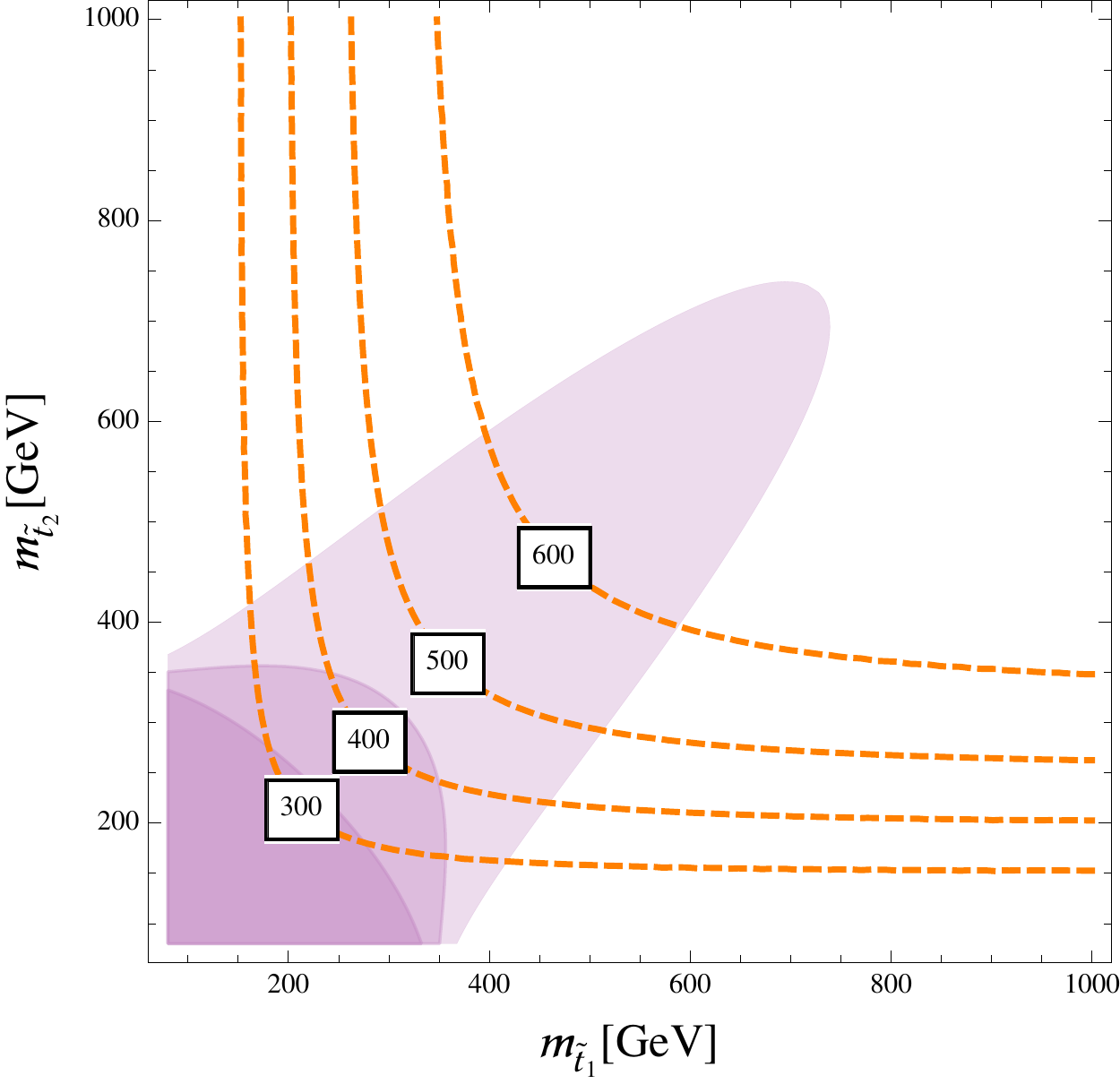} 
\end{center}
\caption{Left: global fit in the $(r_G^{\tilde t}, m_H)$ plane. The dotted, dashed, and solid purple contours denote the boundaries of the allowed region at $1 \sigma$, $2 \sigma$, $3 \sigma$ C.L. Right: contours of $m_H$ that minimizes local $\chi^2$ (orange dashed lines) in the stop mass plane. The three shaded purple regions, from darkest to lightest, are excluded at $3 \sigma$ (99.73\%) level; $2 \sigma$ (95.45\%) level; and $1 \sigma$ (68.27\%) level.  }
\label{fig:Dterm}
\end{figure}%

Now we want to estimate the size of the deviation in $r_b$ in a concrete model. As discussed in~\cite{Blum:2012ii}, in the simplest non-decoupling $D$-term models (e.g.~\cite{Batra:2003nj, Maloney:2004rc}) without a hard PQ-symmetry breaking source, 
\beq
r_b &\approx  &\left(1-\frac{m_h^2}{m_H^2}\right)^{-2} \left(1-\frac{\delta^{\tilde{t}}_{m_h^2}}{m_H^2}\right)\approx\left(1-\frac{m_h^2}{m_H^2}\right)^{-2}, \nonumber \\
&\approx &1 + 0.22 \left(\frac{400 \,{\rm GeV}}{m_H}\right)^2
\label{eq:rbd}
\eeq
where $m_H$ is the heavy CP-even Higgs mass. $\delta^{\tilde{t}}_{m_h^2}$ is the stop loop contribution to $m_h^2$ and $\left(1-\frac{\delta^{\tilde{t}}_{m_h^2}}{m_H^2}\right)$ will only introduce a correction $\lesssim$ 5\% for $m_H \gtrsim 400$ GeV. Thus it is a good approximation that in these models, $r_b$ is only determined by $m_H$. We replot the fit in the $(r_G^{\tilde t}, m_H)$ plane in the left panel of Fig.~\ref{fig:Dterm}. We also plot contours of $m_H$ that minimizes local $\chi^2$ in the stop mass plane in the right panel of Fig.~\ref{fig:Dterm}. From the figure, one could see that to allow a larger positive deviation in $r_G^{\tilde{t}}$ from two light stops, the heavy Higgs has to be light and within the reach of direct searches (or indirect constraints, of which there are many~\cite{Altmannshofer:2012ks}). More specifically, if the experimental heavy Higgs mass bound is pushed to be 500 GeV or above, the enhancement in the bottom Yukawa is not sufficient to compensate a possible increase in the Higgs digluon coupling. Effectively, the two parameter fit will be reduced to one parameter fit with a stronger exclusion on the stop masses in Sec.~\ref{sec:case1}.

\subsection{Case 3: Higgs Mixing Effect at Small $\tan\beta$}
\label{sec:case3}
SUSY models that rely on an additional tree-level $F$-term contribution to raise the light Higgs mass above $m_Z$ mostly work in the small $\tan\beta$ region. Typical models include a new singlet in the Higgs sector with examples such as $\lambda$SUSY~\cite{Gherghetta:2012gb,Hall:2011aa}, the NMSSM~\cite{Espinosa:1991gr, Ellwanger:2009dp}, and its variant the DiracNMSSM~\cite{Lu:2013cta}. Strictly speaking, the Higgs sector is no longer 2HDM if the singlet is light and one should include one more parameter quantifying the singlet--Higgs mixing in addition to $r_b$ and $\tan\beta$~\cite{D'Agnolo:2012mj}. In this section, we consider a simper case where the new singlet is heavy and could be integrated out leaving the low energy Higgs sector still as a 2HDM. This could be realized, e.g., in the DiracNMSSM~\cite{Lu:2013cta}. Our purpose is to test the robustness of the Higgs coupling constraints on the stop masses derived in the previous two cases and find the conditions under which these constraints could be relaxed. 

In this case, there are four parameters relevant for the Higgs data fit: $r_G^{\tilde t}, r_\gamma^{\tilde \chi}, r_b, \tan \beta$. $r_\gamma^{\tilde \chi}$, the contribution from charginos to the Higgs diphoton coupling, is included as it might be non-negligible when $\tan \beta$ is close to 1. We restrict $\tan \beta \geq 1$ and $r_\gamma^{\tilde \chi}$ to be in the range depicted in Fig. 1 of ~\cite{Blum:2012ii} derived taking all charginos to be heavier than 94 GeV~\cite{Beringer:1900zz}.

\begin{figure}[!h]\begin{center}
\includegraphics[width=0.43\textwidth]{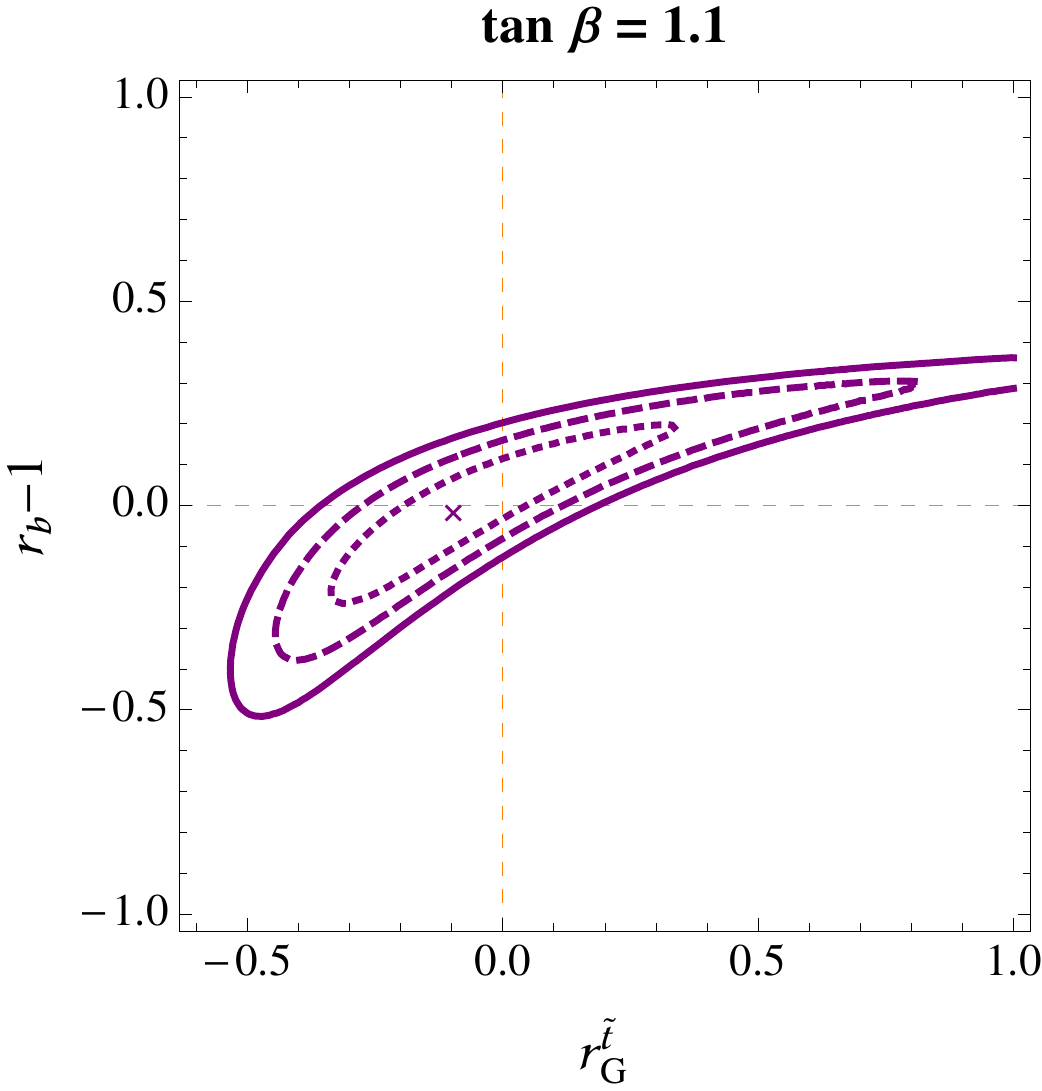}  \quad \includegraphics[width=0.43\textwidth]{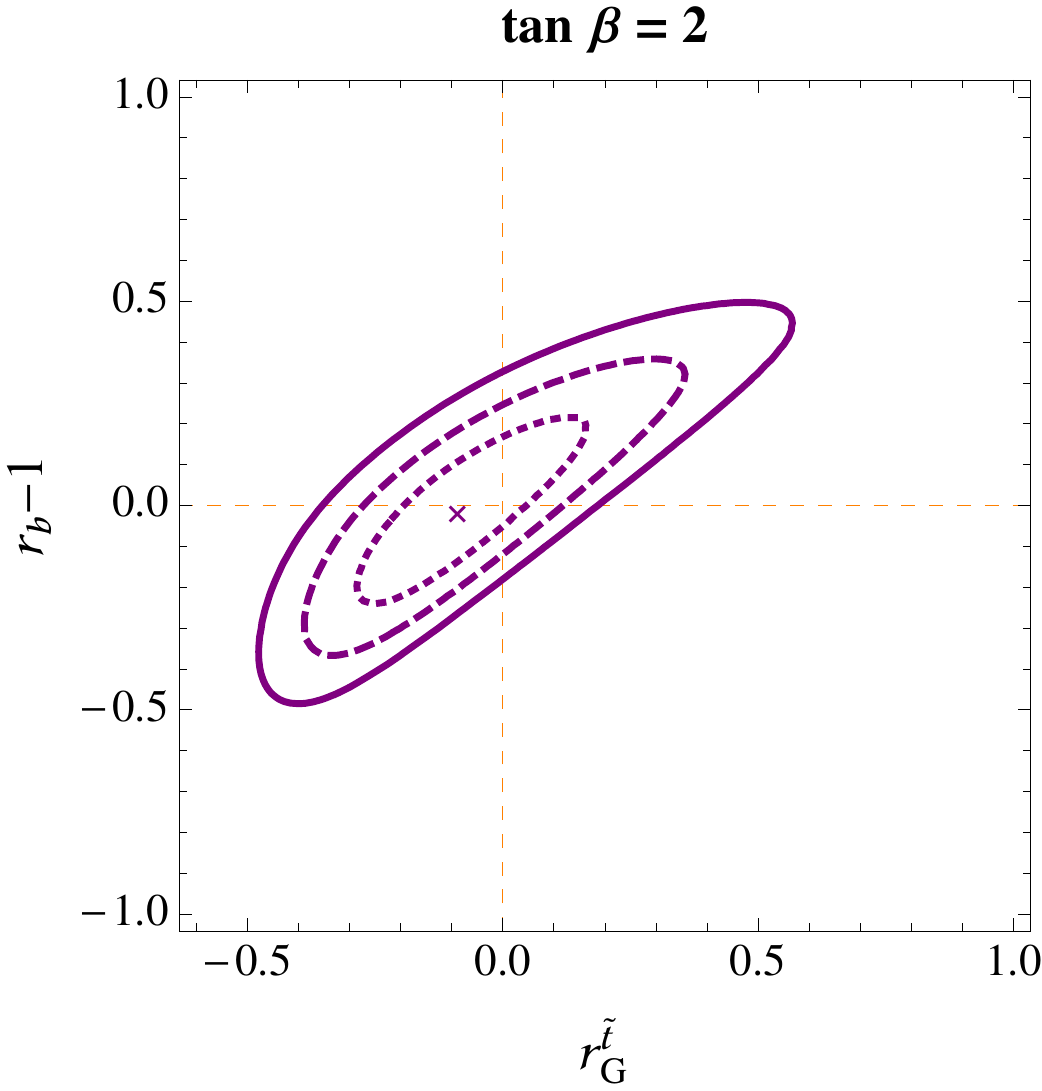}
\end{center}
\caption{Global fits in the $(r_G^{\tilde t}, r_b-1)$ plane assuming $\tan \beta = 1.1$ (left) and $\tan \beta =2$ (right). $\times$ denotes the best fit point. The dotted, dashed, and solid purple contours denote the boundaries of the allowed region at $1 \sigma$, $2 \sigma$, $3 \sigma$ C.L. }
\label{fig:fitvaryingtb}
\end{figure}%

\begin{figure}[!h]\begin{center}
\includegraphics[width=0.45\textwidth]{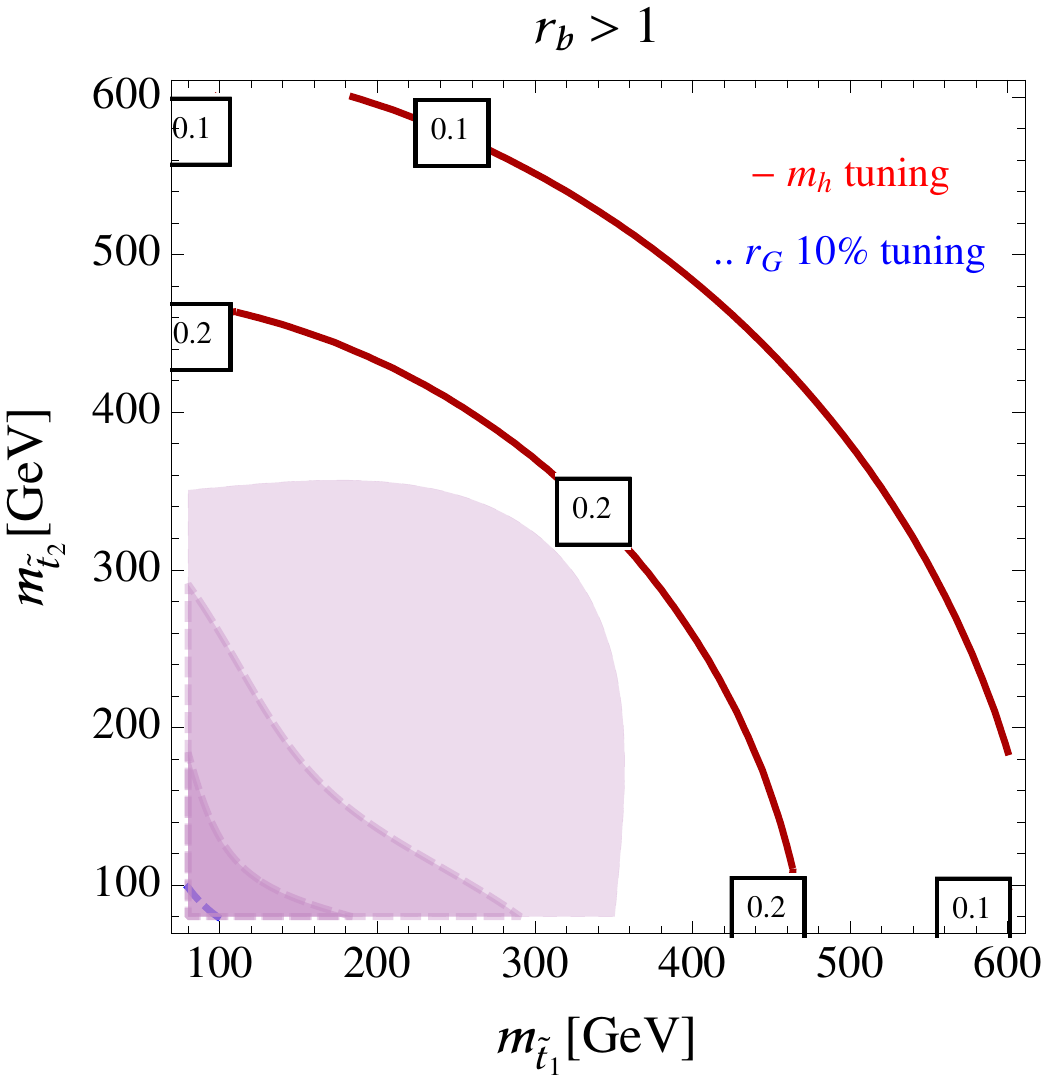} 
\end{center}
\caption{Assuming no other contributions to Higgs digluon coupling $r_G$ other than stops', region of natural stop that has been ruled out by Higgs coupling measurements with varying $r_b, \tan \beta$ and $r_\gamma^{\tilde{\chi}}$. The three shaded purple regions, from darkest to lightest, are excluded at $3 \sigma$ (99.73\%) level; $2 \sigma$ (95.45\%) level; and $1 \sigma$ (68.27\%) level.  The red solid lines: contours of Higgs mass fine-tuning. The fit prefers small $\tan\beta$ and hence larger $y_t$, which contributes to the tuning. Blue dashed lines: contour of 10\% fine-tuning associated with $r_G^{\tilde{t}}$. Note that the constraints are only valid for models with $r_b > 1$ and $\tan \beta$ close to 1. For scenarios with $r_b < 1$ or $\tan \beta \geq 3$, the constraints are the same as those demonstrated in Sec.~\ref{sec:case1} or Sec.~\ref{sec:case2}. Detailed explanations are in the text.  }
\label{fig:higgscoupling3}
\end{figure}%

We first study the effects of $\tan\beta$ on the fit. The results are demonstrated in Figure~\ref{fig:fitvaryingtb}. We pick two values of $\tan\beta$ and for each point in $(r_G^{\tilde t}, r_b-1)$ plane, vary $r_\gamma^{\tilde \chi}$ to minimize $\chi^2$. When $\tan\beta$ is close to 1, an enhancement in $r_b$ is associated with a considerable reduction in $r_t$ and correspondingly $r_G$, which is 
\beq
r_G = r_t (r_G^{\tilde{t}}+1).
\eeq
 Thus the stop loop contribution $r_G^{\tilde{t}}$ is allowed to take larger positive values in the fit. When $\tan\beta$ gets larger, the effect of modification of $r_b$ on $r_t$ decreases. Thus the fit becomes less sensitive to $\tan\beta$. When $\tan\beta \geq 3$, the allowed region in $(r_G^{\tilde t}, r_b-1)$ plane is always the same as depicted in Fig.~\ref{fig:fitp2} as the fit is effectively only sensitive to two parameters $r_G^{\tilde{t}}, r_b$. We also find that the fit is not very sensitive to the variation of $r_\gamma^{\tilde \chi}$ in the theoretically allowed range. 
 
Finally we perform a profile likelihood fit to map out the allowed region in the plane of the stops' physical masses. This is depicted in Fig.~\ref{fig:higgscoupling3}. Now the constraints are greatly relaxed compared to those derived in the previous two cases. This is mostly due to the anti-correlation between $r_b$ and $r_t$ at $\tan \beta =1$. If $\tan\beta \geq 3$, the fit is reduced to the two parameter fit and the constraints will be the same as that in the second case. Besides, the relaxation of the constraints is only achieved with a positive $r_b-1$ and consequently a negative $r_t-1$. However, in the models such as DiracNMSSM, the bottom Yukawa coupling is always reduced compared to its SM value. In that case, the Higgs coupling constraints on the stop masses are the same as the strong ones depicted in Fig.~\ref{fig:higgscoupling}! In the general NMSSM with a light singlet mixing with the Higgs, the bottom Yukawa coupling might be enhanced if the heavy Higgs is light~\cite{Blum:2012ii}. In that case, the Higgs coupling constraints on the stop masses could be ameliorated. 

\subsection{Prospects from LHC Run 2 and Future Colliders}
\label{sec:future}

We have seen that measurements of Higgs properties have already begun to constrain a large part of the stop parameter space for which fine-tuning is less than around a factor of ten. It is interesting to ask what the prospects are for improving these constraints in the near future with LHC Run 2, as well as from possible future colliders like the ILC or TLEP that would perform precision measurements of Higgs properties. We have used the Snowmass Higgs working group estimates~\cite{Dawson:2013bba} to perform a simple estimate of this reach. We assume that the Higgs has Standard Model couplings which are measured to be 1 with an error bar given by Table 1-20 of ref.~\cite{Dawson:2013bba}, and examine the one-parameter fit for $r_G^{\tilde t}$ given these constraints. For instance, the table lists a precision of 2 to 5\% on the photon coupling and 3 to 5\% on the gluon coupling at the high luminosity LHC. We take the center of these ranges and assume the couplings are measured to be $1 \pm 0.035$ and $1 \pm 0.04$ times their SM values, then examine what range of stop parameter space would be excluded.

\begin{figure}[!h]\begin{center}
\includegraphics[width=1.0\textwidth]{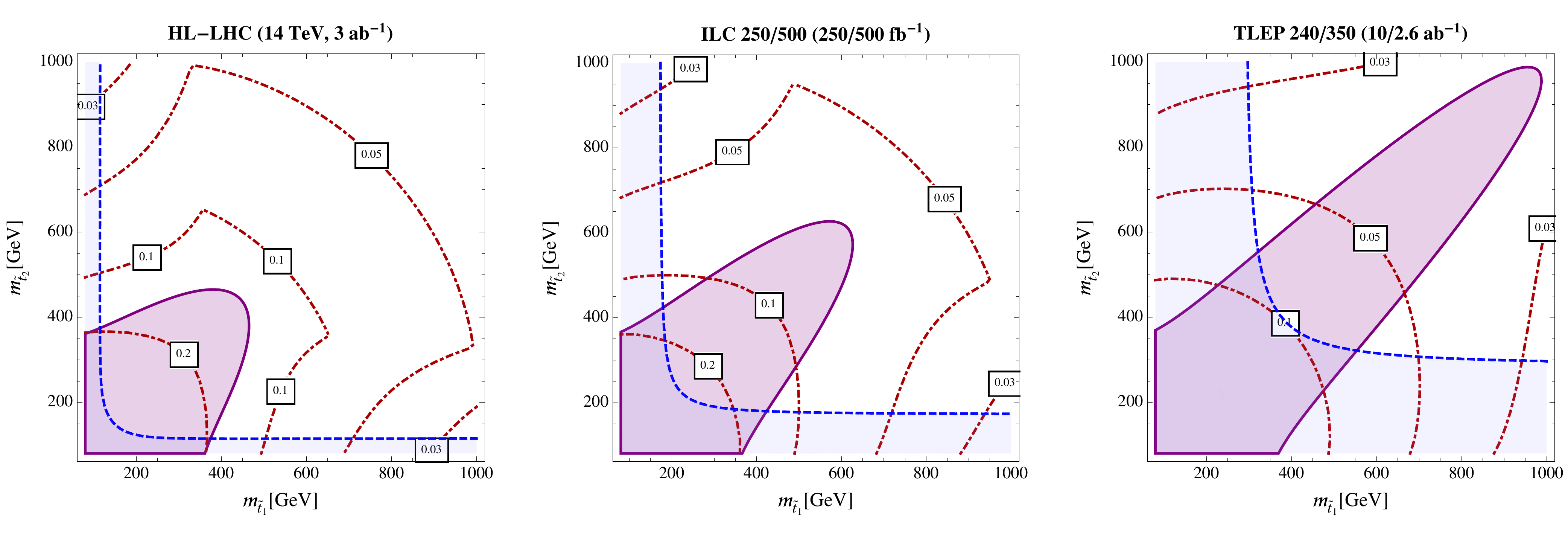} 
\end{center}
\caption{Projected constraints on stops (from a one-parameter fit) from future experiments. The purple shaded region along the diagonal has a minimum $\left|X_t\right|$ needed to fit the data at 95\% CL that is larger than $\left|X_t^{\rm max}\right|$. The blue shaded region requires a tuning of $X_t$ by more than a factor of 10 to fit the data. The dot-dashed red contours label Higgs mass fine-tuning.}
\label{fig:futureprojection}
\end{figure}%

In Figure~\ref{fig:futureprojection}, we show the resulting projected reach of three experiments: the high luminosity LHC Run 2 assuming 3 ab$^{-1}$ of data; the ILC running at 250 and 500 GeV and collecting 250 and 500 fb$^{-1}$ of data; and TLEP running at 240 and 350 GeV and collecting 10 and 2.6 ab$^{-1}$ of data. Notice that the HL-LHC projection is no better than the current exclusion in Fig.~\ref{fig:higgscoupling}. This indicates that we have been ``lucky'' so far, in the sense that current data prefers a decreased Higgs coupling to gluons, and we have a stronger exclusion than expected. As precision increases at the ILC or TLEP, the constraint from the $\left|X_t^{\rm min}\right| > \left|X_t^{\rm max}\right|$ argument extends along the diagonal, ruling out nearly-degenerate stops up to high masses as precision increases. However, as discussed in Section~\ref{sec:basic}, the exclusion region is anchored at 350 GeV on both axes, and we see that the constraint does not extend far from the diagonal. As the precision of the measurements increases, the exclusion based on tuning of Higgs couplings becomes progressively more important, as indicated by the shaded blue regions in the figure. Furthermore, because the value of $\left|X_t^{\rm min}\right|$ for given stop masses increases with the precision of the measurement and $A_t$ enters the tuning measure, we can see that the tuning curves move inward over time. TLEP would completely rule out regions of 10\% tuning, as well as a slice of parameter space with even higher fine-tuning. The ILC or TLEP would also directly constrain higgsinos, and thus pin down tree-level fine-tuning as well as the loop effects we discuss.

\section{Constraints on Folded Stops}
\label{sec:folded}

In light of our failure to find supersymmetry so far, one could wonder if naturalness of electroweak symmetry breaking might be enforced by a more subtle mechanism. One such theoretical proposal is Folded Supersymmetry~\cite{Burdman:2006tz}, in which top partners still cancel loop corrections to the Higgs mass, but these top partners have no Standard Model SU(3)$_c$ quantum numbers. However, these ``$F$-stops'' still have electroweak quantum numbers, which are necessary to allow them to couple to the Higgs boson. They would contribute loop corrections to the $h \to \gamma\gamma$ amplitude but not to the $h \to gg$ amplitude. The Higgs also acquires a new decay to hidden gluons, $h \to g_h g_h$, which may or may not appear as an invisible width experimentally depending on the lifetime of the hidden sector glueballs, but in any case is very small and does not affect the fits. Because the $W$ loop dominates over the top loop in the SM $h \to \gamma\gamma$ amplitude, the loop corrections from $F$-stops are more difficult to observe than those of ordinary stops (which show up dominantly in the coupling $h \to gg$). Still, we can ask how well the LHC and future colliders can constrain $F$-stops, and whether measurements of the $h \to \gamma\gamma$ amplitude could be complementary to studies of Higgs wavefunction renormalization as a probe of naturalness in this scenario~\cite{Craig:2013xia}.

\begin{figure}[!h]\begin{center}
\includegraphics[width=0.8\textwidth]{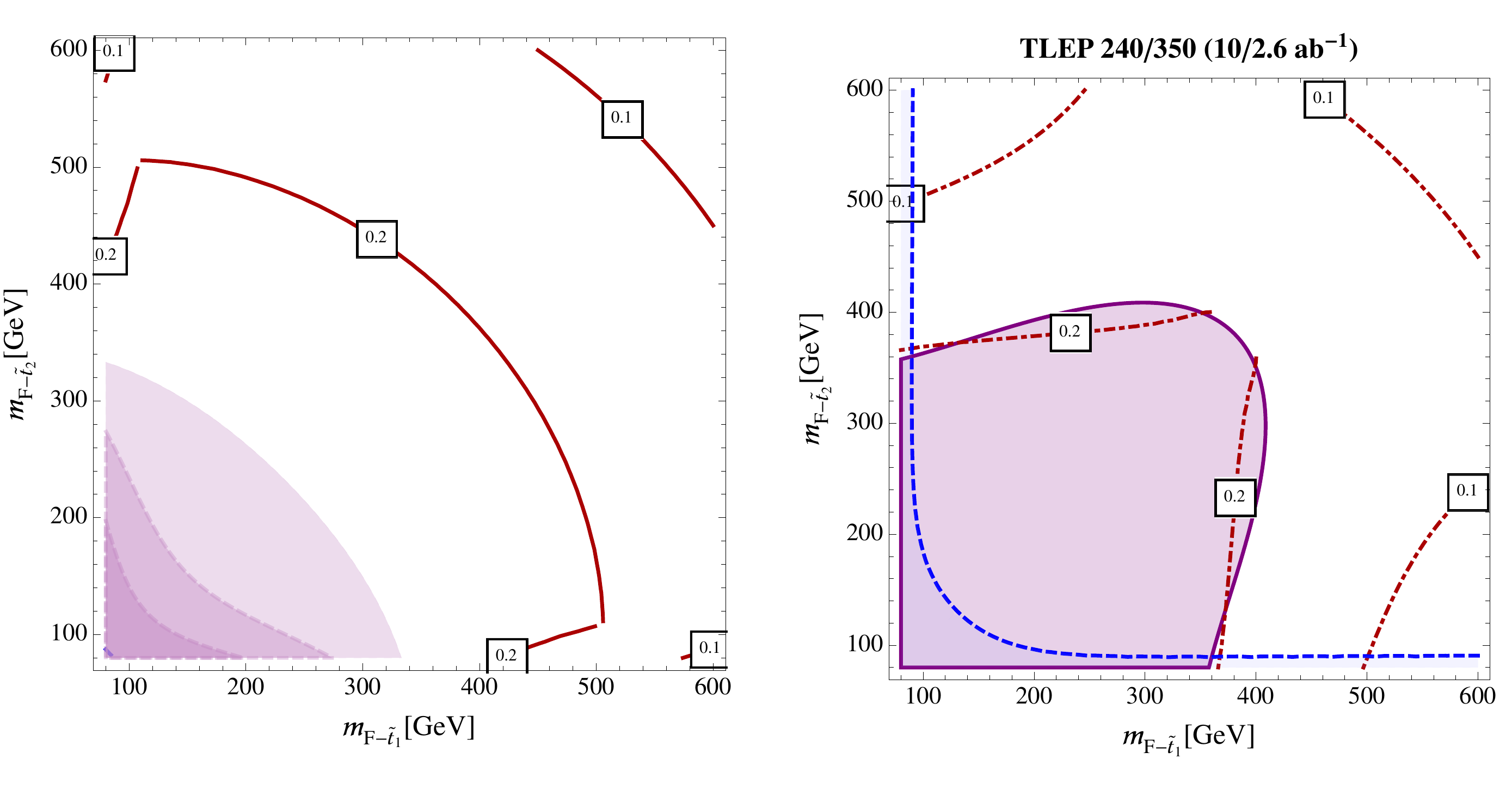} 
\end{center}
\caption{Constraints on folded SUSY-like models from a one-parameter fit with ``$F$-stops,'' i.e. scalars that have the electroweak quantum numbers and Yukawa couplings of stops but no QCD charge. Because they affect $h \to \gamma\gamma$ but not $h \to gg$, constraints on such particles are much weaker than on stops. At left are the current constraints. The three shaded purple regions, from darkest to lightest, are excluded at $3 \sigma$ (99.73\%) level; $2 \sigma$ (95.45\%) level; and $1 \sigma$ (68.27\%) level.  The red solid lines: contours of Higgs mass fine-tuning. The blue dashed line displaying 10\% fine-tuning associated with $r_\gamma^{\tilde{t}}$ is barely visible in the left-hand corner, indicating that we do not yet have enough precision to make this argument. At right: projected constraints from TLEP. The purple shaded region along the diagonal has a minimum $\left|X_t\right|$ needed to fit the data at 95\% CL that is larger than $\left|X_t^{\rm max}\right|$. The blue shaded region requires a tuning of $X_t$ by more than a factor of 10 to fit the data. The dot-dashed red contours label Higgs mass fine-tuning.}
\label{fig:higgscouplingfold}
\end{figure}%

The original model of Folded SUSY makes fairly specific predictions for the mass spectrum, but here we just assume the existence of $F$-stops that have all of the couplings of ordinary stops except for the coupling to gluons. The constraints arising from the $F$-stops' modifications of the $h \to \gamma\gamma$ decay width are plotted in Fig.~\ref{fig:higgscouplingfold}, which also shows projected TLEP reach. These constraints are significantly weaker than constraints on ordinary stops, reinforcing the idea that ``colorless supersymmetry'' is a challenging scenario to constrain with the LHC. Even a future collider like TLEP, which would set very powerful constraints on ordinary stops, would only constrain folded stops to about the 20\% tuning level. (Other colorless supersymmetry scenarios also typically involve new electroweak states that might alter Higgs properties; see, for instance, refs.~\cite{Chang:2006ra,Craig:2013fga}.)

\section{Discussion}
\label{sec:discussion}

\subsection{Possible Caveats}
\label{sec:outlook}
In our analysis, we neglect beyond-MSSM physics in loops, assuming that the leading loop correction to the Higgs-gluon coupling originates from stops. If there exist light vector-like colored states beyond MSSM which contribute negatively to the Higgs-gluon coupling, the constraints on stop masses might be relaxed. However, the cancelation between the new colored states and stops still contributes to the Higgs coupling fine-tuning. 

When including Higgs mixing effects in Sec.~\ref{sec:case2} and Sec.~\ref{sec:case3}, we neglect non-holomorphic bottom and tau Yukawa couplings that could arise from integrating out third generation squarks, higgsinos and gauginos at the one-loop level. Such non-holomorphic Yukawas would alter the 2HDM coupling relations that we assumed. They are only non-negligible when $\tan\beta$ is large, e.g, $\tan \beta \gtrsim 50$. However SUSY scenarios with such a large $\tan\beta$ are always fine-tuned at worse than 1\% level in flavor observables such as $B_s \to X_s \gamma$~\cite{Blum:2012ii, Altmannshofer:2012ks}. Thus we do not consider these scenarios here further. 

\subsection{What is Tuning?}
\label{sec:tuning}

Attitudes about fine-tuning vary widely in the particle theory community. We have seen in Figure~\ref{fig:higgscoupling} that at 95\% confidence level, theories where the dominant corrections to Higgs properties arise from stop loops are constrained to be tuned at worse than the 20\% level (according to the measure in equation~\ref{eq:Dz}). A 10\% fine-tuning is still compatible with the data at 90\% confidence level, although a substantial portion of the parameter space with less than 10\% tuning is already ruled out.
Theorists often discuss models that are much more tuned, so one might wonder how significant this result is. We believe it is an important conclusion. Of course, to some extent this is an aesthetic judgment, and in any case it relies on intuitions about the structure of the space of UV completions of the Standard Model that are hard to make precise and rigorous. Nonetheless, we will briefly respond to a few of the objections that have arisen to taking such a tuning argument seriously.

{\bf Is the QCD scale even more tuned than $m_Z$?} The origin of eq.~\ref{eq:Dz} for the measure of tuning arising from stop loops is ultimately the Barbieri-Giudice measure~\cite{Barbieri:1987fn}, which measures the tuning of an observable quantity ${\cal O}$ with respect to an underlying parameter $p$ via:
\beq
\left(\Delta_{\cal O}^{-1}\right)_p = \left|\frac{\partial \log {\cal O}}{\partial \log p}\right|.
\eeq
The intuition here is that we look at how volumes in observable space scale as we look at increasingly larger volumes around a point in parameter space. The Barbieri-Giudice measure is fundamentally a measure of the {\em sensitivity} of an observable to an underlying parameter. If we use this definition, we find that the QCD scale exhibits a high degree of sensitivity with respect to the underlying parameter $g$:
\beq
\left(\Delta^{-1}_{\Lambda_{\rm QCD}}\right)_g = \left|\frac{g}{\Lambda_{\rm QCD}} \frac{\partial}{\partial g} \left(\Lambda_{\rm UV} e^{-8\pi^2/(b g^2)}\right)\right| = 2 \log \frac{\Lambda_{\rm UV}}{\Lambda_{\rm QCD}},
\eeq
which is a number $\sim 10^2$, if we take the UV scale to be the GUT or Planck scale. We don't normally view the smallness of the QCD scale compared to these fundamental scales to be a naturalness problem. Why, then, should we be worried about a comparatively small factor of $\sim 10$ sensitivity of the $Z$ mass to the stop masses?

The reason is that the Barbieri-Giudice measure is {\em not} really measuring what most of us think of as tuning. Indeed, most theorists who worry about naturalness rely on the Potter Stewart tuning measure~\cite{Stewart:1964}: we know it when we see it, and the QCD scale is not it. Dimensional transmutation naturally generates small scales. It is {\em sensitive} to the value of $g$ at the high scale, but sensitivity is a weaker statement than tuning. Typically, when we consider a theory to be tuned there must be a {\em cancelation} between numbers of the same order, as when the stop loop correction to $m_{H_u}^2$ cancels against its tree-level value to produce a small number at the weak scale. The amount of this cancelation happens to be captured by the Barbieri-Giudice measure in this case, so we use it as a proxy for the amount of tuning. Indeed, in general if we find that two apparently unrelated numbers cancel to high precision, $\left|x + y\right| \ll \left|x\right|,\left|y\right|$, we might define an intuitive tuning as $\Delta^{-1} \sim \frac{|x|+|y|}{|x+y|}$. In the case of the Higgs corrections, this would be essentially the same thing as $2 \delta m_{H_u}^2/m_h^2$, which is the outcome of the Barbieri-Giudice measure anyway. On the other hand, as QCD illustrates, a theory with a large Barbieri-Giudice tuning measure is not necessarily tuned.

{\bf Do $\lambda$SUSY-like theories escape the argument?} A second point that arises is that, in some theories, there are large tree-level terms beyond those arising in the MSSM, and they modify the usual tuning measure. An example that received much recent attention is $\lambda$SUSY, which reduces the naive amount of tuning by a factor $g^2/\lambda^2$~\cite{Barbieri:2006bg,Hall:2011aa,Gherghetta:2012gb}. However, this is not a panacea. The new large couplings tend to make the Higgs heavier than 125 GeV, and a new source of tuning arises from the need to adjust the singlet mixing to bring the mass back down. Depending on how one measures tuning, this may or may not return us to the measure we started with. In any case, measurements of the Higgs properties disfavor the large values of $\lambda$ for which this objection applies~\cite{Farina:2013fsa}.

{\bf Can UV physics rescue us from tuning?} The most subtle counterargument is that computing tuning based on low scale masses misses the possibility that the UV theory involves correlations among parameters so that what looks, from the IR point of view, like an accidental cancelation was in fact mandated by the structure of the theory. This argument usually arises in the context of Focus Point SUSY~\cite{Feng:1999mn,Feng:1999zg,Feng:2011aa,Feng:2012jfa,Baer:2013gva}. The basic point here is that if the ultraviolet theory has a set of parameters $p$, we want to compute tuning by studying how the low-energy observables vary with the choice of $p$. When we use an estimate like equation~\ref{eq:Dz}, we are instead studying how low-energy observables vary with Lagrangian parameters (like the stop soft masses) defined at a low scale, which may relate to the underlying parameters in a complicated way through RGEs. It is certainly a true statement that these two tuning measures can differ. But we expect that they will rarely differ in a very significant way. The Focus Point argument relies on a cancelation that robustly happens as a universal scalar mass $m_0^2$ is varied, but which would not happen if $y_t$ took a different value. Thus, if $y_t$ is included as a UV parameter, the argument becomes significantly weaker. It could be that $y_t$ is more rigid than $m_0^2$, but this requires a strong assumption about the space of UV theories (e.g., about the string landscape). Even without varying $y_t$, realistic versions of this scenario are tuned at the part-in-a-hundred to part-in-a-thousand level~\cite{Draper:2013cka}.

None of this discussion, of course, implies that supersymmetry is unlikely to be realized in our universe. What it does imply, at least from our point of view, is that any realization of supersymmetry in our universe is likely to look fine-tuned from the standpoint of low-energy effective field theory, by at least a factor of ten. Continuing experimental searches for new particles and for deviations in Higgs properties are needed to strengthen or overturn this conclusion. We emphasize that the scenarios with weaker constraints on stops required the Higgs to mix with other light Higgs bosons, so searches for such heavy Higgses should be a key part of the LHC strategy to constrain naturalness.

\section*{Acknowledgments}
We thank Tim Cohen, Maxim Perelstein, and Josh Ruderman for discussions.

\appendix

\section{Tuning with RG-induced $A_t$}\label{app:AtermRG}

\begin{figure}[!h]\begin{center}
\includegraphics[width=0.45\textwidth]{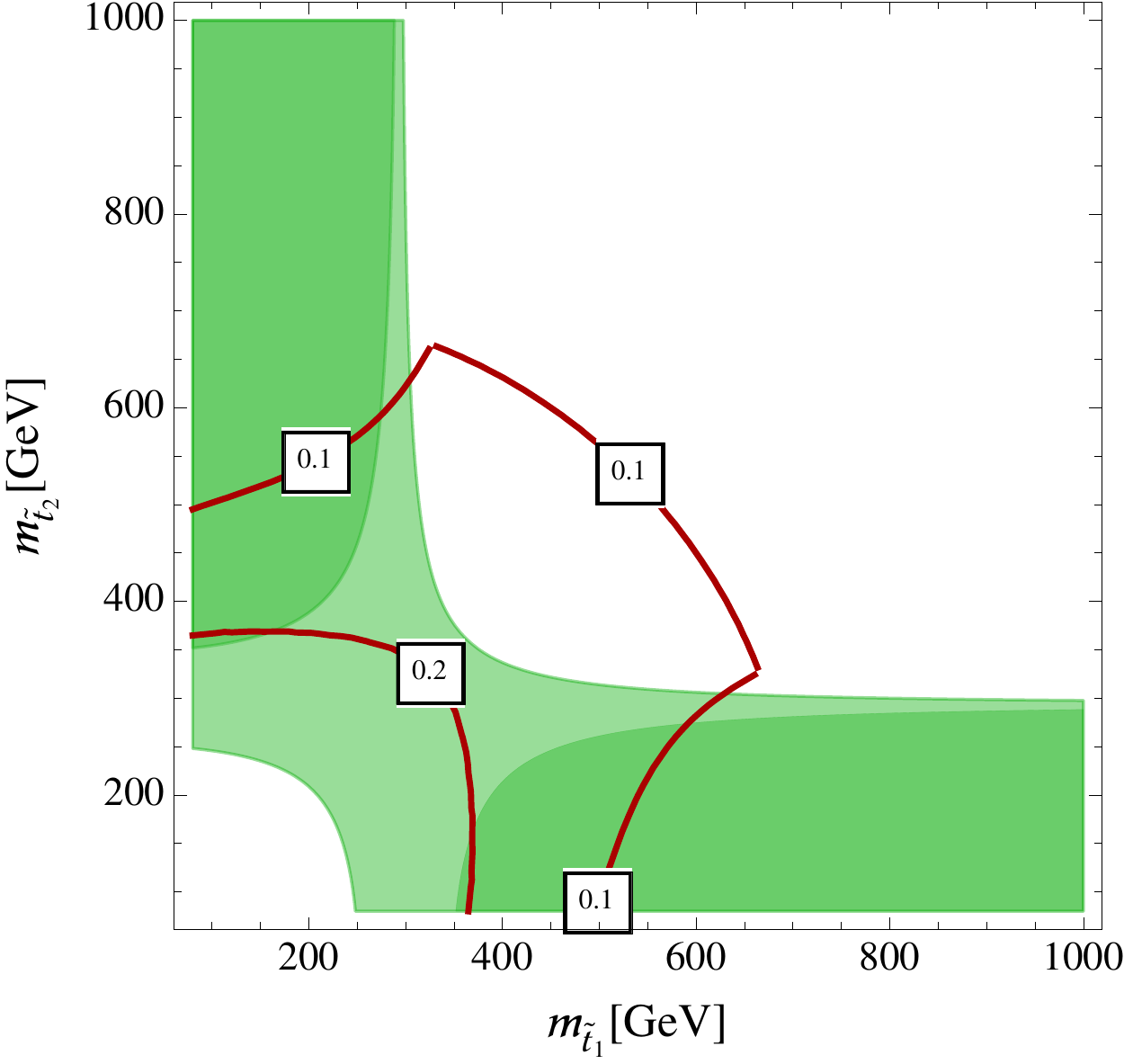}
\end{center}
\caption{Regions in which, if $A_t$ is generated from the gluino mass through RGEs, the gluino mass contributes tuning of a factor of 10 (darker green) or a factor of 5 (lighter green) to the Higgs mass. The red contours are as in Fig.~\ref{fig:higgscoupling}, i.e. they consider only the stop contribution to the tuning.}
\label{fig:boundAtfromM3}
\end{figure}%

Many models generate $A$-terms dominantly through RGE effects, though exceptions exist~\cite{Craig:2012xp}. In models where the $A$-term is induced from the gluino mass $M_3$ through loops, we can make a further inference about tuning. The one-loop RG-induced $A$-term is
\beq
A_t \approx -\frac{2}{3\pi^2} g_3^2 M_3 \log\left(\frac{\Lambda}{M_3}\right),
\eeq
while the gluino contributes to tuning at two loops according to~\cite{Papucci:2011wy}
\beq
\left.\delta m^2_{H_u}\right|_{\rm gluino} \approx - \frac{2}{\pi^2} \frac{y_t \alpha_s}{\pi} M_3^2 \log^2\left(\frac{\Lambda}{M_3}\right).
\eeq
As a result, the tuning $\left(\Delta_h^{-1}\right)_{\tilde g}$ is directly related to the smallest $A_t$ consistent with the data:
\beq
\left(\Delta_h^{-1}\right)_{\tilde g} \gsim \frac{9 y_t}{4 g_3^2} \left(\frac{A_t^{\rm min}}{m_h}\right)^2.
\eeq

In Figure~\ref{fig:boundAtfromM3} we show the regions in which the gluino contribution from fine-tuning is large, if the $A_t$ value needed to fit the data arises from the gluino mass through RGs. As in Section~\ref{sec:constraints}, we take $A_t = \left|X_t^{\rm min} \right|+\mu/\tan \beta$ with the SUSY breaking mediation scale $\Lambda = 30$ TeV, $\mu = -200$ GeV and $\tan\beta =10$. In this sort of scenario where $A_t$ arises dominantly from RGEs, we see that the data favor {\em both} stops being heavier than around 300 GeV.

\section{Higgs Data}\label{app:data}
In this appendix, we list all the channels of Higgs data we used in the fit in Table~\ref{tab:CMS}, \ref{tab:atlas}, \ref{tab:tevatron}. 
\begin{table}[!hhh]
\footnotesize
\centering
\renewcommand{\arraystretch}{1.1}
\begin{tabular}{|l | c | c | c |}
\hline
Channel & $\mu$ & $\zeta_i^{\rm (G,V,T)}$ (\%) & Refs. \\
\hline\hline 
$b \bar b$ (VBF) & ${1.0\pm 0.5}$ & $(0,100,0)$ &~\cite{Chatrchyan:2013zna} \\
\hline
$\tau \bar{\tau}\,({0j})$ & ${0.34 \pm 1.09}$ & $(98.1, 1.9,0)$ & \\
$\tau \bar{\tau}\,({1j})$ & ${1.07 \pm 0.46}$ & $(77.3, 22.7, 0)$ &~\cite{cmstau} \\
$\tau \bar{\tau}\,({2j})$ & ${0.94 \pm 0.41}$ & $(19.0, 81.0,0)$ & \\
$\tau \bar{\tau}\,({VH})$ & ${-0.33 \pm 1.02}$ & $(0,100,0)$ &\\
\hline
$WW\, (0/1j)$ & ${0.74^{+0.22}_{-0.2}}$ & $(95.7, 4.3, 0)$ &    \\
$WW\, (2j;$ VBF) & ${0.6^{+0.57}_{-0.46}}$ &$(22.3, 77.7,0)$ &~\cite{Chatrchyan:2013iaa} \\
$WW\, (3 \ell 3 \nu)$ & ${0.56^{+1.27}_{-0.95}}$ & $(0,100,0)$ &\\
\hline
$ZZ\, (0/1 j)$ & $0.83^{+0.31}_{-0.25}$ & $(92.8, 7.2, 0)$ & \\
$ZZ\, (2j)$ & $1.45^{+0.89}_{-0.62}$ & $(54.8, 42.5 , 2.7)$ &~\cite{Chatrchyan:2013mxa} \\
\hline
$\gamma \gamma $ (untagged 0; 7 TeV)  & $3.78^{+2.01}_{-1.62}$ & $(61.4, 35.5,3.1)$& \\
$\gamma \gamma $ (untagged 0; 8 TeV)  & $2.12^{+0.92}_{-0.78}$& $(72.9,24.6,2.6)$ & \\
$\gamma \gamma $ (untagged 1; 7 TeV) & $0.15^{+0.99}_{-0.92}$ & $(87.6,11.8,0.5)$ & \\
$\gamma \gamma $ (untagged 1; 8 TeV) &$-0.03^{+0.71}_{-0.64}$ & $(83.5,15.5,1.0)$ &  \\
$\gamma \gamma $ (untagged 2; 7 TeV) & $-0.05 \pm 1.21$& $(91.3,8.3,0.3)$ & \\
$\gamma \gamma $ (untagged 2; 8 TeV) &$0.22^{+0.46}_{-0.42}$ & $(91.7,7.9,0.4)$ &  \\
$\gamma \gamma $ (untagged 3; 7 TeV) & $1.38^{+1.66}_{-1.55}$& $(91.3,8.5,0.2)$ &~\cite{cmsgamma12, cmsgamma13} \\
$\gamma \gamma $ (untagged 3; 8 TeV) & $-0.81^{+0.85}_{-0.42}$ & $(92.5,7.2,0.2)$ &\\
$\gamma \gamma $ (dijet; 8 TeV) & $4.13^{+2.33}_{-1.76}$& $(26.8,73.1,0.0)$ &  \\
$\gamma \gamma $ (dijet loose; 8 TeV) &  $0.75^{+1.06}_{-0.99}$ & $(46.8,52.8,0.5)$ & \\
$\gamma \gamma $ (dijet tight; 8 TeV) & $0.22^{+0.71}_{-0.57}$ & $(20.7,79.2,0.1)$ & \\
$\gamma \gamma $ (MET; 8 TeV) & $1.84^{+2.65}_{-2.26}$ & $(0.0,79.3,20.8)$ & \\
$\gamma \gamma $ (Electron; 8 TeV)  & $-0.70^{+2.75}_{-1.94}$ &  $(1.1,79.3,19.7)$ & \\
$\gamma \gamma $ (Muon; 8 TeV)  & $0.36^{+1.84}_{-1.38}$ &  $(21.1,67.0,11.8)$ & \\
\hline
\end{tabular}
\caption{\small CMS data used in fits.  Official values for efficiencies are used when quoted, otherwise approximations are made according to a channel's primary topologies. Unless specificed, the signal strengths are derived from a combination of 7 TeV and 8 TeV data. $\zeta_i^{\rm (G,V,T)}$ stand for weights of gluon fusion channel ($G$), vector boson fusion plus associated production with $W, Z$ channels ($V$) and associated production with tops channel ($T$).  }
\label{tab:CMS}
\end{table}

\begin{table}[!hhh]
\footnotesize
\centering
\renewcommand{\arraystretch}{1.1}
\begin{tabular}{|l | c | c | c |}
\hline
Channel & $\mu$ & $\zeta_i^{\rm (G,V,T)}$ (\%) & Refs. \\
\hline\hline 
$b \bar b$ & ${0.2^{+0.7}_{-0.6}}$ & $(0,100,0)$ &~\cite{atlasb} \\
\hline
$\tau \bar{\tau}$ (boosted) & ${1.2^{+0.8}_{-0.6}}$ & $(66, 34,0)$ & \\
$\tau \bar{\tau}$ (VBF) & ${1.6^{+0.6}_{-0.5}}$ & $(10, 90, 0)$ &~\cite{atlastau} \\
\hline
$WW\, (0/1j)$ & ${0.82^{+0.33}_{-0.32}}$ & $(98, 2, 0)$ &    \\
$WW\, (2 j)$ & ${1.4^{+0.7}_{-0.6}}$ & $(19,81,0)$ &~\cite{Aad:2013wqa}\\
\hline
$ZZ$ (other) & $1.45^{+0.43}_{-0.36}$ & $(90.4, 9.6, 0)$ & \\
$ZZ$ (VBF + VH)& $1.2^{+1.6}_{-0.9}$ & $(37.0, 63.0 , 0)$ &~\cite{Aad:2013wqa} \\
\hline
$\gamma \gamma $ (low $p_T$) & $1.6^{+0.5}_{-0.4}$ & $(91.1, 8.6, 0.3)$& \\
$\gamma \gamma $  (high $p_T$)& $1.7^{+0.7}_{-0.6}$& $(78.6, 19.9, 1.4)$ & \\
$\gamma \gamma $  ($2j$)& $1.9^{+0.8}_{-0.6}$ & $(32.3,67.7,0)$ & \\
$\gamma \gamma $ (VH)&$1.3^{+1.2}_{-1.1}$ & $(22.4,68.1,9.5)$ &~\cite{Aad:2013wqa}\  \\
\hline
\end{tabular}
\caption{\small ATLAS data used in fits.  Official values for efficiencies are used when quoted, otherwise approximations are made according to a channel's primary topologies.}
\label{tab:atlas}
\end{table}

\begin{table}[!hhh]
\footnotesize
\centering
\renewcommand{\arraystretch}{1.1}
\begin{tabular}{|l | c | c | c |}
\hline
Channel & $\mu$ & $\zeta_i^{\rm (G,V,T)}$ (\%) & Refs. \\
\hline\hline 
$b \bar b$ & ${1.59^{+0.69}_{-0.72}}$ & $(0,100,0)$ &~\cite{Aaltonen:2013kxa} \\
\hline
$\tau \bar{\tau}$ & ${1.68^{+2.28}_{-1.68}}$ & $(50, 50,0)$ &~\cite{Aaltonen:2013kxa} \\
\hline
$WW$ & ${0.94^{+0.85}_{-0.83}}$ & $(77.5, 22.5, 0)$ &~\cite{Aaltonen:2013kxa}    \\
\hline
$\gamma \gamma $ & $5.97^{+3.39}_{-3.12}$ & $(77.5, 22.5, 0)$&~\cite{Aaltonen:2013kxa} \\
\hline
\end{tabular}
\caption{\small Combined CDF and D0 data used in fits.  Efficiencies for $\tau\tau$ channel are approximated from~\cite{Abazov:2012ee}.}
\label{tab:tevatron}
\end{table}

\section{HiggsSignals Comparison}\label{app:HScompare}

We have used the HiggsSignals 1.1.0 software~\cite{Bechtle:2013xfa,Bechtle:2008jh,Bechtle:2013gu,Bechtle:2013wla} as a check that our own fits produce reasonable results. At this time, HiggsSignals does not include some of the latest experimental updates that appeared after October 2013 (e.g. the ATLAS tau update~\cite{atlastau}). On the other hand, HiggsSignals is a well-tested code with a more thorough treatment of uncertainties than our simple $\chi^2$ fit. The HiggsSignals example code HSeffC.f90 is easily modified to perform fits analogous to those in Section~\ref{sec:constraints}. 

\begin{figure}[!h]\begin{center}
\includegraphics[width=0.8\textwidth]{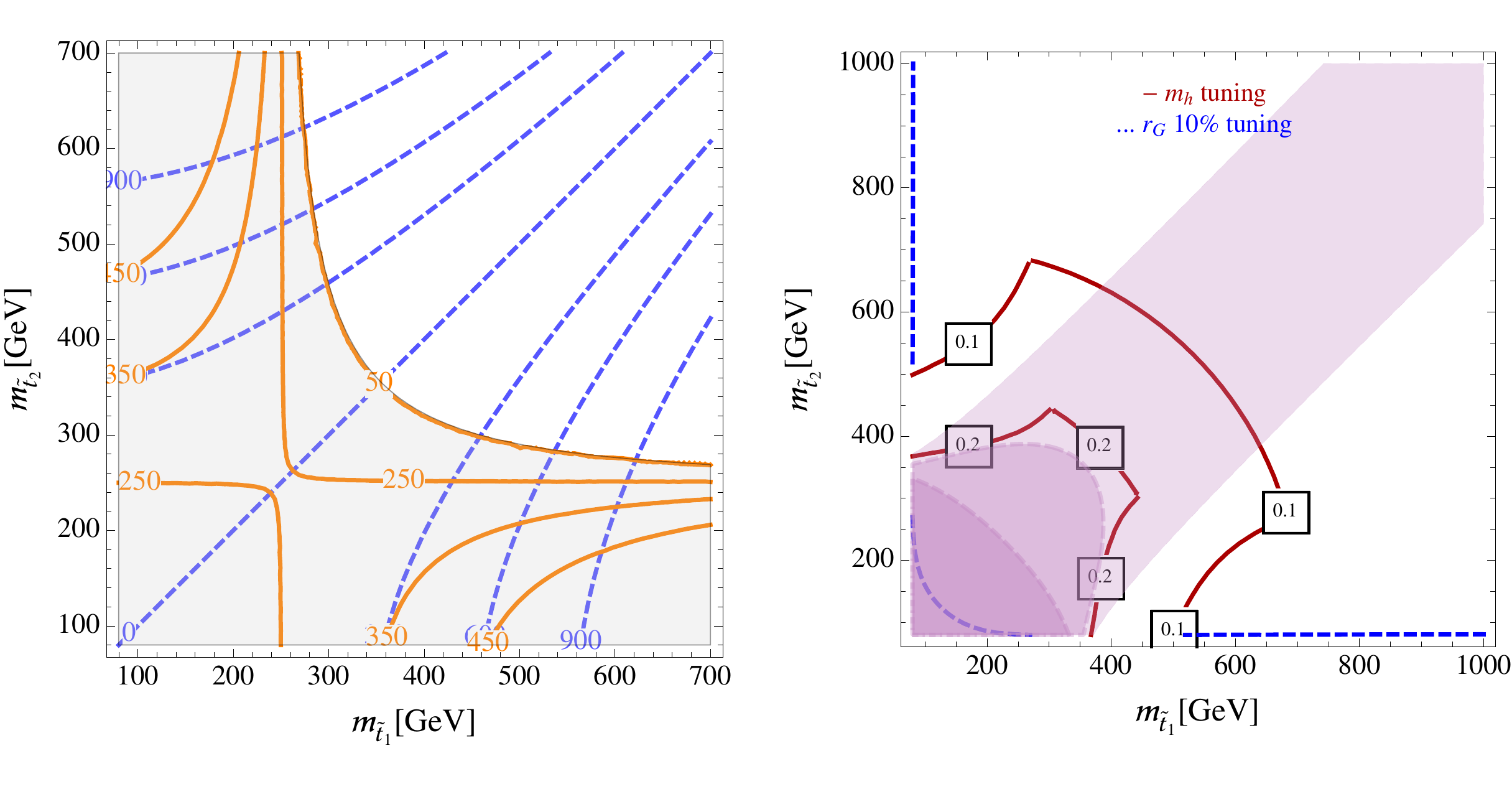}
\end{center}
\caption{Result of a one-dimensional fit using the HiggsSignals code, to be compared with Figures~\ref{fig:methodillustrate} and~\ref{fig:higgscoupling}. The results are similar, but the 2$\sigma$ limits extracted with HiggsSignals are somewhat weaker (reaching slightly below 400 GeV along the diagonal, rather than slightly above). The difference is partly due to our use of more recent ATLAS and CMS updates, and partly due to a different treatment of uncertainties.}
\label{fig:HScompare}
\end{figure}%

We present the results of the one-dimensional fit using HiggsSignals in Figure~\ref{fig:HScompare}. The HiggsSignals result gives somewhat (but not dramatically) weaker $2\sigma$ and $3\sigma$ exclusion contours. We have checked that part of the difference arises from our use of updated data, but part is due to a different treatment of uncertainties. The lesson we take from this is that our fits give reasonably good results, but a full treatment of all correlated uncertainties by the experimental collaboration would be welcome to allow a sharper statement of the current exclusion.

\bibliography{ref}
\bibliographystyle{utphys}
\end{document}